%% file: mnras_template.tex
\documentclass[fleqn,usenatbib]{mnras}

\usepackage{newtxtext,newtxmath}

\usepackage[T1]{fontenc}

\DeclareRobustCommand{\VAN}[3]{#2}
\let\VANthebibliography\thebibliography
\def\thebibliography{\DeclareRobustCommand{\VAN}[3]{##3}\VANthebibliography}

\usepackage{graphicx}	
\usepackage{amsmath}	
\usepackage{caption}
\usepackage{subcaption}
\usepackage{color}
\usepackage{xcolor}
\usepackage{ulem}

\input{shortcuts}

\title[The Columba project]{Columba: isolated dwarf galaxy populations in diverse cosmological environments simulated with a cold interstellar medium}

\author[J. M. Briggs et al.]{Jemima M. Briggs,$^{1,2}$\thanks{E-mail: J.M.Briggs@2016.ljmu.ac.uk}
Azadeh Fattahi,$^{2,3}$
Robert A. Crain,$^{1}$
Yannick M. Bahe,$^{4,5}$
Sylvia Ploeckinger,$^{6}$
\newauthor
Matthieu Schaller,$^{7,8}$
Alejandro Ben\'itez-Llambay,$^{9}$
Evgenii Chaikin,$^{8}$
Alexander J. Richings,$^{10,11}$
\newauthor
and Joop Schaye$^{8}$
\\
$^{1}$Astrophysics Research Institute, Liverpool John Moores University, 146 Brownlow Hill, Liverpool, L3 5RF\\
$^{2}$Institute for Computational Cosmology, Department of Physics, University of Durham, South Road, Durham DH1 3LE\\
$^{3}$The Oskar Klein Centre, Department of Physics, Stockholm University, Albanova University Center, 106 91 Stockholm, Sweden\\
$^{4}$School of Physics and Astronomy, University of Nottingham, University Park, Nottingham NG7 2RD\\
$^{5}$Institute of Physics, Ecole Polytechnique Fédérale de Lausanne (EPFL), Observatoire de Sauverny, 1290 Versoix, Switzerland\\
$^{6}$Department of Astrophysics, University of Vienna, Türkenschanzstrasse 17, 1180 Vienna, Austria\\
$^{7}$Lorentz Institute for Theoretical Physics, Leiden University, PO Box 9506, 2300 RA Leiden, the Netherlands\\
$^{8}$Leiden Observatory, Leiden University, PO Box 9513, 2300 RA Leiden, the Netherlands\\
$^{9}$Dipartimento di Fisica G. Occhialini, Università degli Studi di Milano Bicocca, Piazza della Scienza, 3 I-20126 Milano MI, Italy\\
$^{10}$Centre for Data Science, Artificial Intelligence and Modelling, University of Hull, Cottingham Road, Hull, HU6 7RX\\
$^{11}$E. A. Milne Centre for Astrophysics, University of Hull, Cottingham Road, Hull, HU6 7RX
}

\date{Accepted XXX. Received YYY; in original form ZZZ}

\pubyear{2015}

\begin{document}
\label{firstpage}
\pagerange{\pageref{firstpage}--\pageref{lastpage}}
\maketitle

\begin{abstract}
\correction{We introduce a suite of $\Lambda$CDM cosmological, hydrodynamical simulations that track the evolution of a large population of dwarf galaxies. The suite comprises zoom-in simulations of 25 spherical, under-dense regions of $r=5\,\cMpc$, selected to span $\approx1.5$ dex in mean enclosed density, covering voids to filamentary structures, whilst excluding haloes of Milky~Way-mass or larger. The simulations achieve a mass resolution of $\sim10^5\Msun$ with a galaxy formation model including cold, dense interstellar gas and whose subgrid stellar feedback efficiency reproduces the $z=0$ galaxy stellar mass function. We investigate the impact of the cosmic environment on dwarf galaxy formation and evolution. We find that the $5\,\cMpc$ environment influences the normalisation of the halo and galaxy mass functions, but does not significantly affect the stellar mass - halo mass (SMHM) relation and halo occupation fraction for galaxies with $M_{\star}=10^6-10^9 \Msun$. Instead, host halo concentration, estimated from DM-only counterparts, is more important: both the fraction of haloes hosting a resolved galaxy and the scatter about the SMHM relation correlate positively with concentration. Owing to halo assembly bias, concentration also influences galaxy formation times, such that at fixed halo mass more concentrated haloes host galaxies that are both older and more massive. The offset from the mean SMHM relation also anti-correlates with $t_{90}$, the time at which 90 percent of a galaxy's stellar mass has assembled. These correlations between halo properties and galaxy star formation histories present testable predictions for forthcoming observational surveys.}
\end{abstract}




\input{Introduction/introduction}
\input{Methods/methods}

\input{Validation/validation}

\input{Results/results}

\input{Summary/summary}

\section*{Acknowledgements}
We thank Adrian Jenkins for generous advice concerning the generation of the initial conditions for this project. JMB and AF acknowledge support by a UK Research and Innovation Future Leaders Fellowship [grant no MR/T042362/1]. AF is supported by a Wallenbeger Academy Fellowship. RAC acknowledges support from STFC Small Astronomy Awards ST/Y002474/1 and ST/Y002482/1. 
SP acknowledges support from the Austrian Science Fund (FWF) through grant number V 982-N. ABL acknowledges support by the Italian Ministry for Universities (MUR) program ‘Dipartimenti di Eccellenza 2023-2027’ within the Centro Bicocca di Cosmologia Quantitativa (BiCoQ), and support by UNIMIB’s Fonuk redo Di Ateneo
Quota Competitiva (project 2024-ATEQC-0050). YMB acknowledges support from UKRI through a Future Leaders Fellowship (grant agreement MR/X035166/1). 
This work used the Prospero HPC facility at Liverpool John Moores University, and the DiRAC@Durham facility managed by the Institute for Computational Cosmology on behalf of the STFC DiRAC HPC Facility (www.dirac.ac.uk). The equipment was funded by BEIS capital funding via STFC capital grants ST/K00042X/1, ST/P002293/1, ST/R002371/1 and ST/S002502/1, Durham University and STFC operations grant ST/R000832/1. DiRAC is part of the National e-Infrastructure.
This research has made use of NASA's Astrophysics Data System. 

\section*{Data Availability}
Data produced in this study can be available upon reasonable request to the authors. 

A public version of the \textsc{Swift} hydrodynamics code is available at \url{http://www.swiftsim.com}. The \textsc{chimes} non-equilibrium thermochemistry code is publicly available at \url{https://richings.bitbucket.io/chimes/home.html}.



\bibliographystyle{mnras}
\bibliography{example} 




\appendix
\input{appendix}


\bsp	
\label{lastpage}
\end{document}

%% file: shortcuts.tex
\newcommand{\correction}{\textcolor{black}}

\newcommand{\Msun}{\rm{M_\odot}}
\newcommand{\Msunperyear}{\rm{M_\odot}\,{\rm yr}^{-1}}
\newcommand{\kpc}{\rm{kpc}}
\newcommand{\ckpc}{\rm{ckpc}}
\newcommand{\pkpc}{\rm{pkpc}}
\newcommand{\Mpc}{\rm{Mpc}}
\newcommand{\cMpc}{\rm{cMpc}}

\newcommand{\hMpc}{h^{-1}\rm{Mpc}}
\newcommand{\kms}{\rm{km\,s}^{-1}}
\newcommand{\Myr}{\rm{Myr}}
\newcommand{\Gyr}{\rm{Gyr}}

\newcommand{\dfive}{\delta_{\rm 5}}
\newcommand{\cnfw}{c_{\rm NFW}}
\newcommand{\cnfwmath}{c_\mathrm{NFW}}
\newcommand{\vmax}{V_{\rm max}}
\newcommand{\mstar}{M_{\star}}
\newcommand{\mhalo}{M_{\rm 200}}

%% file: Introduction/introduction.tex
\section{Introduction}

Understanding the formation and evolution of dwarf galaxies is a problem at the frontiers of both cosmology and astrophysics. A number of outstanding challenges to the prevailing $\Lambda$-Cold Dark Matter ($\Lambda$CDM) cosmogony on dwarf galaxy scales are still debated in the literature, such as the the `missing satellites' \citep{Klypin_1999,Moore_1999}, `cusp-core' \citep[e.g.][]{Flores_1994} and `too big to fail' \citep{Boylan-Kolchin_2011} problems \citep[see][for a review]{Bullock_BoylanKolchin_2017}. These open questions are often cited as an indication that the $\Lambda$CDM paradigm requires revision or extension, but their solution may also follow from an improved understanding of how dwarf galaxies form and evolve, and the concomitant influence of these processes on the structure of dwarf galaxy dark matter (DM) haloes \citep[see e.g.][]{Sales_2022}.

Dwarf galaxies are of particular interest as `laboratories' for studying the astrophysics of galaxies, as their early formation times and shallow gravitational potentials render them particularly sensitive to the physical processes that regulate galaxy growth. These include both internal mechanisms such as radiative and mechanical feedback from young stars and supernovae (SNe), and external influences such as the UV/X-ray ionising background radiation or interactions with a more massive host galaxy. Supernova feedback has long been shown to play a particularly important role in removing baryons from low mass haloes and hence reconciling the steep mass function of dark matter haloes with the shallower galaxy stellar mass function \citep[GSMF, e.g.][]{Dekel1986,Lacey1991,Bower2012} in the dwarf galaxy regime. The UV/X-ray ionising background radiation is expected to evaporate gas from low-mass haloes (those with maximum circular velocity, $\vmax \lesssim 30 \kms$) leaving a significant fraction of them devoid of baryons at the present day \citep{Quinn1996,Shapiro2004,Crain2007} and quenching star formation in the small fraction of haloes of this scale that were able to initiate star formation prior to the epoch of reionisation \citep[e.g.][]{Ricotti2005}. 

Almost all ultra faint dwarf galaxies (those with $\mstar \lesssim 10^5 \Msun$) are believed to be relics of the early Universe, having formed almost entirely prior to the epoch of reionisation \citep[see][for a review]{Simon2019}. More massive dwarf galaxies are believed to `reignite' after reionisation \citep{BenitezLlambay_2015}. These predictions are consistent, at face value, with observations of dwarf galaxies in and around the Local Group. Indeed, ultra faint dwarf galaxies are mostly old and metal poor, whereas brighter dwarf galaxies exhibit diverse star formation histories \citep{Weisz2011,Weisz2014,Gallart_2015}.

The sensitivity of Local Group dwarf galaxies to their environment is apparent in the dichotomy of the morphology of dwarf satellite galaxies and more isolated counterparts. The majority of the satellites of the Milky Way and M31 are gas-poor dwarf spheroidal galaxies, but the isolated dwarf galaxies beyond the immediate environment ($r \gtrsim 300\,{\rm kpc}$) of the Milky Way and M31 exhibit significant neutral hydrogen reservoirs and are star forming \citep[see][and references therein]{McConnachie2012}. These trends extend to galaxies beyond the Local Group. \citet{Geha_2012} find a correlation between the fraction of quenched dwarf galaxies and the distance to the nearest massive galaxy using SDSS data, finding that isolated dwarf galaxies in the mass range $10^7\Msun<\mathit{M}_{\star}<10^9 \Msun$ are seldom quenched\footnote{The study stops at lower mass end of $\mstar \sim 10^{7} \Msun$ due to incompleteness of data.}. On the other hand, satellite dwarf galaxies around MW-like hosts are efficiently quenched, as shown by recent results from the SAGA survey \citep{Geha2024}.

Aside from the immediate neighbourhood, the environment defined on larger scales has also been shown to influence properties of galaxies and DM haloes. The clustering of DM haloes depends on secondary halo properties such as their concentration and formation time, in particular for low mass haloes, with more concentrated and older low mass haloes more likely to be found in denser environments, as demonstrated in simulations \citep{Gao_2005,Wechsler_2006}. Similarly, \citet{Hellwing_2021} predict that low mass ($\mhalo \lesssim 10^{11} \Msun$) haloes in voids exhibit systematically lower concentrations than is typical at fixed mass. The observable properties of galaxies also have been shown to depend on the larger scale environment. For example, \citet{Zehavi_2011} find that red galaxies in the SDSS redshift survey are more clustered, i.e. live in higher density environments. Additionally, semi-analytical and hydrodynamical simulations show that the overdensity in which a halo resides (on $\simeq 5\,Mpc$ scales) correlates with the residual of the stellar mass-halo mass relation about the mean \citep{Zehavi2018, Artale2018}.  

The degree to which environmentally-driven biases of DM haloes influence the observable properties of dwarf galaxies remains poorly understood, due to both observational and theoretical challenges. The majority of known low-mass dwarf galaxies are located in the Local Group with the completeness dropping sharply beyond $\sim 3$Mpc. The structural properties of dwarf galaxies beyond the Local Group are also poorly constrained owing to their intrinsic faintness. Recent attempts to extend the observed dwarf galaxy sample, such as the SAGA \citep{Geha2017, Mao2021} or ELVES \citep{Carlsten2022} surveys, have focused on satellites of MW-like systems, where interaction with the host may significantly influence their properties and hinder investigation of the role of other internal or external physical processes. These observational hurdles have resulted in limited understanding of dwarf galaxies as a population, and even fundamental scaling relations, such as the average stellar mass as a function of halo mass, are poorly constrained below $\mstar < 10^8 \Msun$ \citep[e.g.][]{Read_2017}. Nevertheless, this situation is rapidly changing. For example, the Euclid Deep Field, DR1, and the Euclid Wide Survey are estimated to identify approximately 9900, 470000 and 2.6 million dwarf galaxy candidates \citep{Marleau_2025}, respectively. These developments make it increasingly important to understand the formation of isolated dwarf galaxies in diverse environments, and to develop theoretical predictions that can be confronted with these new datasets.

Cosmological hydrodynamical simulations have proven a powerful means of obtaining insight into the mechanisms that govern the formation and evolution of dwarf galaxies, offering plausible solutions to some of the small-scale tensions faced by the $\Lambda$CDM cosmogony \citep{Sales_2022}. Simulations of representative cosmic volumes (side length $L \sim $ 100 comoving Mpc, hereafter cMpc) have, over the last decade, matured such that they yield galaxy populations with average properties that broadly correspond to those of the observed population \citep[see e.g.][]{Crain_van_de_Voort_2023}. However, the numerical study of dwarf galaxies necessitates high resolution in order to capture the physical mechanisms that govern their evolution and to sample their structure and star formation history. This unavoidably restricts simulations to small volumes and therefore limits the size of simulated dwarf galaxies samples and the diversity of the simulated environments they are found in.

To achieve high resolution in simulations at a reasonable computational cost, the zoom-in method is commonly adopted \citep[e.g.][]{Katz_1993}, wherein the immediate environment of an individual galaxy is followed with gas dynamics at high resolution, and the remaining cosmological volume is evolved with collisionless dynamics at reduced resolution to ensure the galaxy experiences the correct tidal forces. This method has been used to resimulate Milky-Way and Local Group-like analogues, including populations of field and satellite dwarf galaxies in these environments, in projects such as (but not limited to) Eris \citep{Guedes_2011}, APOSTLE \citep{Sawala_2016}, Auriga \citep{Grand_2017}, ARTEMIS \citep{Font_2020}, the D.C. Justice League Simulations \citep{Applebaum_2021}, NIHAO-UHD \citep{Buck_2020}, Latte \citep{Wetzel_2016} and ELVIS \citep{Garrison-Kimmel_2014}. Zoom simulations have also been widely used to follow individual dwarf galaxies in isolation at extremely high resolution, enabling ultra-faint dwarf galaxies to be simulated e.g. the EDGE project \citep{Rey_2019,Rey_2020}, LYRA \citep{Gutcke_2021}, and dwarf galaxies from FIRE \citep{Wheeler_2019}. 

However, due to their small volume, zoom simulations of individual dwarf galaxies are ill-suited to the study of the influence of diverse cosmic environments. This limitation is best overcome with zoom simulations of regions that are significantly larger than individual galaxies, without incurring an infeasible computational cost and memory footprint. This approach was adopted for the general galaxy population by the GIMIC simulations \citep{Crain_2009}, which follow the evolution of galaxies within five roughly spherical regions of comoving radius $r=18-25\,h^{-1}\cMpc$ drawn from the $L=500\,h^{-1}\cMpc$ Millennium Simulation volume. The regions were chosen to have overdensities of $(-2,-1,0,+1,+2)$ times the root-mean-square deviation from the mean overdensity, and thus span $\simeq 95$ per cent of the overdensity range in the parent volume whilst simulating only $0.13$ per cent of its volume. A similar approach was adopted by the FLARES project \citep{Lovell_2021} to examine diverse environments in simulations of the epoch of reionisation, and by the MARVEL-ous project to study dwarf galaxies in the environment defined by a cosmic `sheet' \citep{Bellovary_2019,Munshi_2021}.

Here we introduce the Columba suite of simulations of galaxy formation in the $\Lambda$CDM cosmogony, which are designed primarily to examine the influence of large-scale cosmic environment on the evolution of isolated dwarf galaxies, i.e. those not in close proximity to larger structures at the present day. We focus on this subset of dwarf galaxies in order to explore the influence of large-scale environment, and the intrinsic properties of host DM haloes, on the evolution of dwarf galaxies, in the absence of tidal and ram pressure forces exerted by nearby neighbouring galaxies. 

Columba adopts a similar approach to GIMIC; the Columba suite comprises zoom simulations of 25 spherical regions of radius $r=5\,\cMpc$ drawn from a large parent volume of $L=400\,\cMpc$. The main motivation for these simulations is studying \textit{isolated} dwarf galaxies in various environments such as voids and filaments. As the simulations focus on isolated dwarf galaxies, these regions do not sample the cosmic density distribution (on the $5\,\cMpc$ scale of the simulated spheres) uniformly, but preferentially sample underdense regions. This approach enables a large sample of isolated dwarf galaxies, found in regions spanning 1.5 decades in enclosed density, to be followed with resolution better than galaxy formation simulations of large ($L \sim 100\,\cMpc$) periodic volumes such as EAGLE \citep{Schaye_2015,Crain_2015,McAlpine_2016}, IllustrisTNG \citep{Pillepich_2018,Nelson_2018,Nelson_2019} or SIMBA \citep{Dave_2019}.

The galaxy formation model used for Columba adopts several subgrid modules developed for the COLIBRE project \citep{Schaye_COLIBRE, Chaikin_COLIBRE}, combined with elements retained from the earlier EAGLE model. Columba therefore shares COLIBRE's treatments for processes such as turbulence-based star formation, pre-supernova feedback, and black hole seeding and growth, but does not include all of the new physics implemented in COLIBRE. In particular, Columba lacks the live dust model, the turbulent diffusion, and AGN jet feedback adopted in COLIBRE, and it retains the nucleosynthesis yields and Type Ia supernova rates from EAGLE. Columba nevertheless follows many of the key processes that regulate the formation of the cold, dense interstellar medium, including shielding from UV radiation, the formation and dissociation of molecules, and the radiative cooling that such molecules foster (specific differences relative to the COLIBRE model are detailed in the next section). Cosmological simulations of volumes larger than the environments of individual galaxies have recently begun to treat the cold ISM phase explicitly, with notable examples being NewHorizon \citep{Dubois_2021} and FIREbox \citep{Feldmann_2023}. Our approach complements these studies with an explicit focus on diversity in cosmological environments, which cannot currently be achieved with simulations of uniform resolution (FIREbox) or a single zoom-in region (NewHorizon).

This study serves as a reference for the simulation suite, and presents a number of key findings concerning the connection between the cosmic environment and dwarf galaxy formation and evolution. These findings are in general most clearly demonstrated with tailored simulations of this type. The paper is structured as follows. We begin by presenting in Section \ref{sec:methods} our methods, including our strategy for selecting the simulated regions, the generation of the initial conditions (ICs), the numerical and subgrid galaxy formation components of the model used to evolve the ICs to the present day, and our techniques for identifying galaxies and their parent haloes. In Section \ref{sec:validation} we confront basic properties of the simulated galaxies with observational measurements to validate our model. We examine the environmental variation of key diagnostics in Section \ref{sec:environment}, and explore the origin of scatter in the stellar mass - halo mass relation in Section \ref{sec:scatter_smhm}. We summarise and discuss our findings in Section \ref{sec:summary}.

%% file: Methods/methods.tex
\section{Methods}
\label{sec:methods}

In this section we introduce our methods for selecting the simulated regions (Section \ref{sec:methods:regions}) and generating the ICs (Section \ref{sec:methods:ICs}), describe the model we use to evolve the ICs to the present day (Sections \ref{sec:methods:grav_hydro} and \ref{sec:methods:subgrid}), and explain our techniques for identifying galaxies and their parent haloes (Section \ref{sec:methods:galaxy_identification}).  

\subsection{Selection of simulated regions}
\label{sec:methods:regions}

The primary aim of our simulation design is to follow the evolution of dwarf galaxies ($M_\star \lesssim 10^{10}\,\Msun$) that are isolated at the present day, and at the same time live in a diverse range of large-scale cosmic environments. We identify such environments at $z=0$ within a parent simulation of $L=400\,\rm{cMpc}$ with periodic boundary conditions and its mass distribution realised with $N=3008^3$ composite particles representing baryons and DM. The simulation adopts the maximum posterior likelihood cosmological parameter values reported by the Dark Energy Survey team from year three data \citep[][their ‘3 × 2pt + All Ext.’ CDM parameters]{Abbott_2022}, for which $\Omega_{\rm m}=0.306$ yielding a particle mass of $m_{\rm dm}=9.22 \times 10^7\,\Msun$. This parent simulation uses the dark matter only initial conditions (ICs) corresponding to the COLIBRE L400m7 volume \citep{Schaye_COLIBRE}, generated with the same cosmology and phases but without the 4x sampling of dark matter particles used for the COLIBRE ICs. Consequently, the DM particle mass differs from the published L400m7 run by a factor of $\approx 4$. The ICs of this simulation were generated following the same procedure as that adopted by \citet[][see their Section 2.4]{schaye2023flamingo} for the FLAMINGO simulations, using the third-order Lagrangian perturbation software \textsc{monofonicIC} \citep{Hahn_2020}, which suppress discreteness noise, coupled to the \textsc{panphasia} \citep{Jenkins_2013} Gaussian random noise field\footnote{The phase descriptor of this volume is [Panph6,L18,(200557,163876,161484),S1,KK1025,CH3518244376,\newline COLIBRE400].}. The large-scale phases were fixed to the mean. As for FLAMINGO, we adopt linear power and transfer functions computed with \textsc{class} \citep{Lesgourgues_2011, Lesgourgues_Tram_2011}, with separate transfer functions computed and applied for baryons, DM and neutrinos. The ICs were created at $z=127$\footnote{Colibre L400m7 ICs are at $z=63$.} and evolved to $z=0$ assuming collisionless dynamics using the open-source cosmology, gravity, hydrodynamics, and galaxy formation software \textsc{Swift} \citep{Schaller2024swift}. The short- and long-range gravitational forces are computed using a 4th-order fast multipole method and a particle-mesh method solved in Fourier space. The influence on large-scale matter clustering due to the non-zero density of massive neutrinos \citep[e.g.][]{Lesgourgues_2006} is modelled with a grid-based linear response method \citep[e.g.][]{AliHaimoud_2013}. Haloes are identified in the parent simulation using the VELOCI\textsc{raptor} structure finder \citep{Elahi_2019}, which is further described in Section~\ref{sec:methods:galaxy_identification}.

We elect to simulate spherical regions of $r=5\,\cMpc$. The precise size of the regions is not critical, but we find that this choice yields a judicious balance between probing the `extremes' of the density distribution populated by isolated dwarf galaxies, whilst ensuring that even the most underdense regions yield a meaningful population of well-sampled galaxies. A related benefit of simulating relatively small volumes (compared to typical periodic simulations) is that they can usually be accommodated within the memory footprint (512 gigabytes to 1 terabyte) of the individual compute nodes that comprise many modern high-performance computing facilities, simplifying parallelism.

\begin{figure}
    \centering
    \includegraphics[width=\columnwidth]{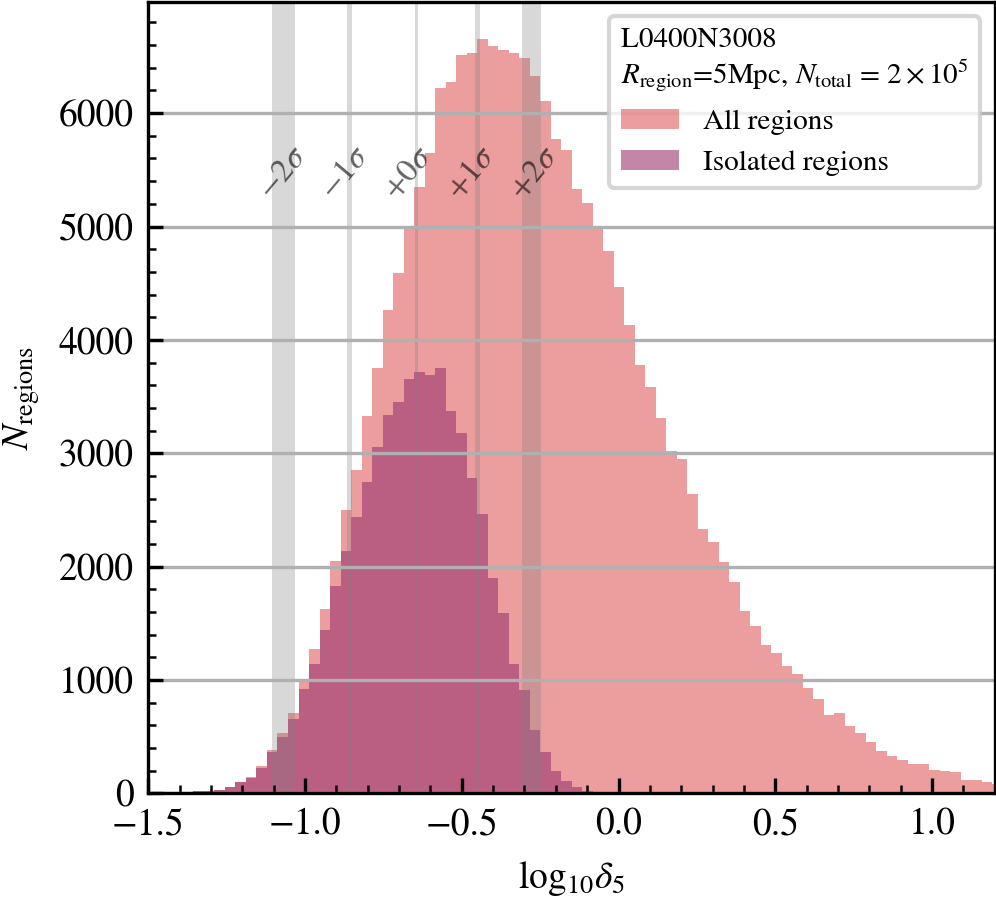}
    \caption{The distribution of mean enclosed density, relative to the cosmic mean density, of $r=5\,\cMpc$ spheres, $\dfive \equiv \rho_5 / \bar{\rho}$, selected from a parent cubic volume of side length $L=400\,\cMpc$, at $z=0$. The background distribution in a lighter shade corresponds to $2\times 10^5$ randomly-centred spheres, the foreground distribution denotes the subset with no haloes of $M_{200} > 5\times10^{11}\,\Msun$ within $r=5\,\cMpc$ of the region centre, and none of mass $M_{200} > 10^{13}\,\Msun$ within $r=10\,\cMpc$. These criteria necessarily bias this subset towards lower density. The vertical bands denote the densities of this subset corresponding to the (-2, -1, 0, +1, +2)$\sigma$ percentile rank ordering of a Gaussian distribution. The width of the bands corresponds to one percentile about these values.}
    \label{fig:overdensity_histogram}
\end{figure}

To select the Columba target regions, we sample the $z=0$ normalised density\footnote{Note that $\delta$ is often used to represent the \textit{overdensity}, i.e. $\delta=(\rho-\bar{\rho})/\bar{\rho}$, but here we use $\delta=\rho/\bar{\rho}$ because the large dynamic range of densities we explore is conveniently expressed logarithmically, requiring that $\delta \geq 0$.}, $\delta_5 = \rho_5 / \bar{\rho}$, where $\rho_5$ is the mean mass density within a sphere of radius $r=5\,\cMpc$ and $\bar{\rho}$ is the mean cosmic mass density, with $2 \times 10^{5}$ spheres randomly-centred within the parent volume. The total sampled volume is roughly twice the volume of the parent box and this ensures that most coordinates within the parent volume are sampled at least once. The resulting density distribution is shown as the lightly-shaded histogram in Fig.~\ref{fig:overdensity_histogram}. Note that this unbiased sample yields, for a sufficiently large sample of regions, a mean density of $\bar{\rho}$, but as a volume-weighted measure the distribution is not symmetric and the median and peak values are less than unity ($\log_{10}\delta_5^{\rm median} = -0.304$, $\log_{10}\delta_5^{\rm peak} \simeq -0.434$).

The darker, foreground histogram shows the density distribution of the subset of regions that are devoid of haloes with mass comparable to or more massive than that of the Milky Way, and which are not in close proximity to a massive halo that would host a group or cluster of galaxies. Specifically, we require that at $z=0$ there are no haloes of mass\footnote{Spherical overdensity masses are defined relative to the critical density, $\rho_{\rm c}$.} $M_{200}>5.0\times 10^{11}\,\Msun$ within $5\,\Mpc$ of the region centre, and no halo of mass $M_{200}>10^{13}\,\Msun$ within $10\,\Mpc$ of the centre. These criteria are satisfied by $5.8\times10^4$ of the initial $2\times10^5$ regions (29 per cent), and are necessarily biased towards lower densities, yielding mean, median and peak normalised densities of $\log_{10} \dfive = -0.653$, $-0.645$ and $-0.568$, respectively. Vertical bands on Fig.~\ref{fig:overdensity_histogram} denote for this sub-sample the densities corresponding to the (-2, -1, 0, +1, +2)$\sigma$ percentile rank ordering of a Gaussian distribution, which have values $\log_{10}\dfive = (-1.064, -0.859, -0.645, -0.449, -0.282)$, respectively. The width of the lines corresponds to 1 percentile. For reference, the same percentile ranks for the full sample (lightly-shaded histogram) correspond to $\log_{10}\dfive = (-0.954, -0.671, -0.304, 0.162, 0.732)$, respectively.

\begin{figure}
    \centering
	\includegraphics[width=\columnwidth]{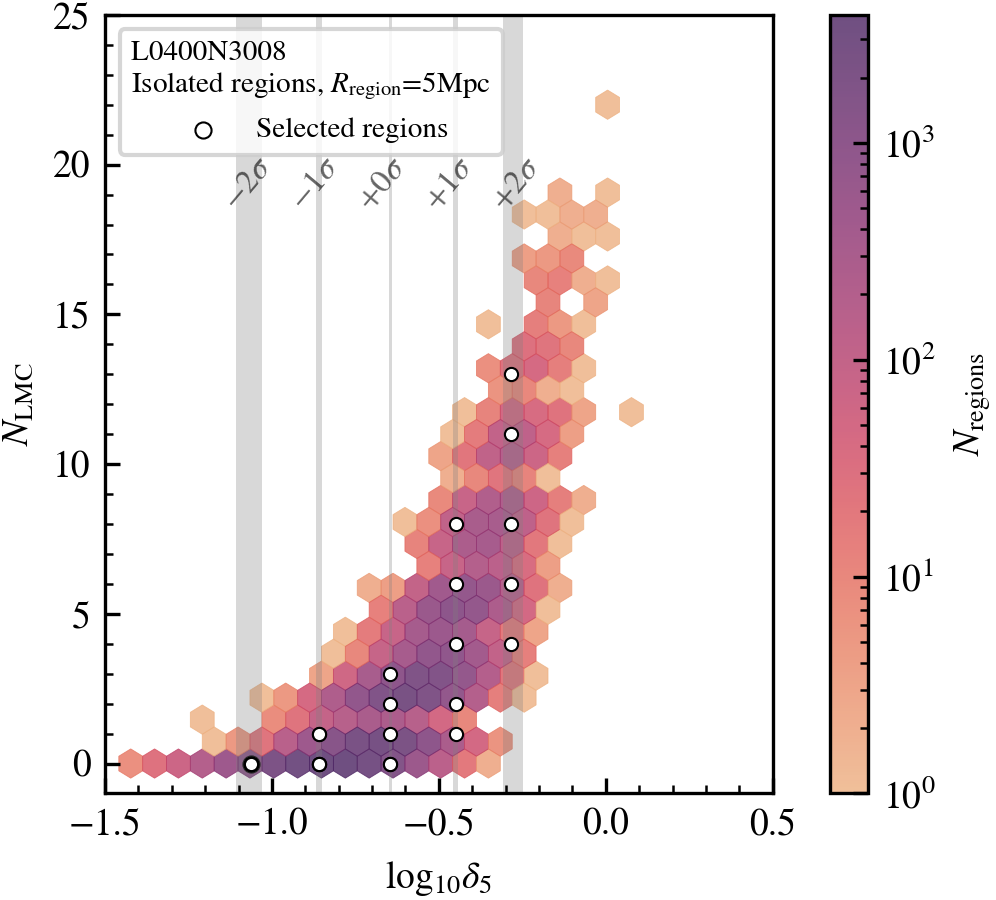}
    \caption{The number of LMC-mass objects ($\mhalo\sim 10^{11} \Msun$) at $z=0$ within the $r=5$Mpc underdense spheres comprising the darker histogram of Fig.~\ref{fig:overdensity_histogram}, as a function of the regions' normalised density, $\dfive$. Hexbins are colour coded by the number of regions. Denser regions tend to host more LMC-mass haloes, but there is significant scatter in $N_{\rm LMC}$ at fixed density. The vertical bands are repeated from Fig. \ref{fig:overdensity_histogram}, and indicate the bands of overdensity in which we select our target regions. The white circles show the values of $N_{\rm LMC}$ that correspond to $(-2, -1, 0, +1, +2)\sigma$ percentiles for the distribution of this quantity within each band. Where fewer than five circles per band are shown, more than one of them overlap. }
    \label{fig:NLMC}
\end{figure}

Despite its inherent bias towards low density regions, the sample of regions with no MW-mass haloes nevertheless exhibits significant diversity in the number of haloes enclosed that have dynamical mass comparable to that of the Large Magellanic Cloud (LMC). Such haloes are amongst the most massive permitted by our selection criteria, so the diversity is necessarily greatest for those regions at the upper end of the density distribution. Fig.~\ref{fig:NLMC} shows a 2-dimensional histogram of the number of regions as a function of enclosed density, $\dfive$, and the number of LMC-mass haloes, $N_{\rm LMC}$, where the latter is defined as $7\times10^{10} \leq M_{200}/\Msun \leq 5\times10^{11}$. $N_{\rm LMC}$ correlates positively with $\dfive$, varying between zero and 22 in the $5.8\times10^4$ regions defining our MW-free sample, but with large scatter overall. We characterise the diversity of $N_{\rm LMC}$ as a function of $\dfive$ for the MW-free sample of regions by computing, for haloes within the 1 percentile-wide density bins denoted by the vertical bands (repeated from Fig.~\ref{fig:overdensity_histogram}), the values corresponding to the Gaussian (-2, -1, 0, +1, +2)$\sigma$ percentile rank ordering of $N_{\rm LMC}$. These values are denoted by white dots overlaid on the vertical bands (where fewer than five dots are visible, two or more of the lowest-$\sigma$ percentiles correspond to $N_{\rm LMC}=0$). For example, in the regions comprising the $+2\sigma$ density sample, the number of LMC-like haloes corresponding to (-2, -1, 0, +1, +2)$\sigma$ are $N_{\rm LMC}=(4, 6, 8, 11, 13)$, respectively.

\begin{figure*}
    \centering
    \includegraphics[width=0.9\textwidth]{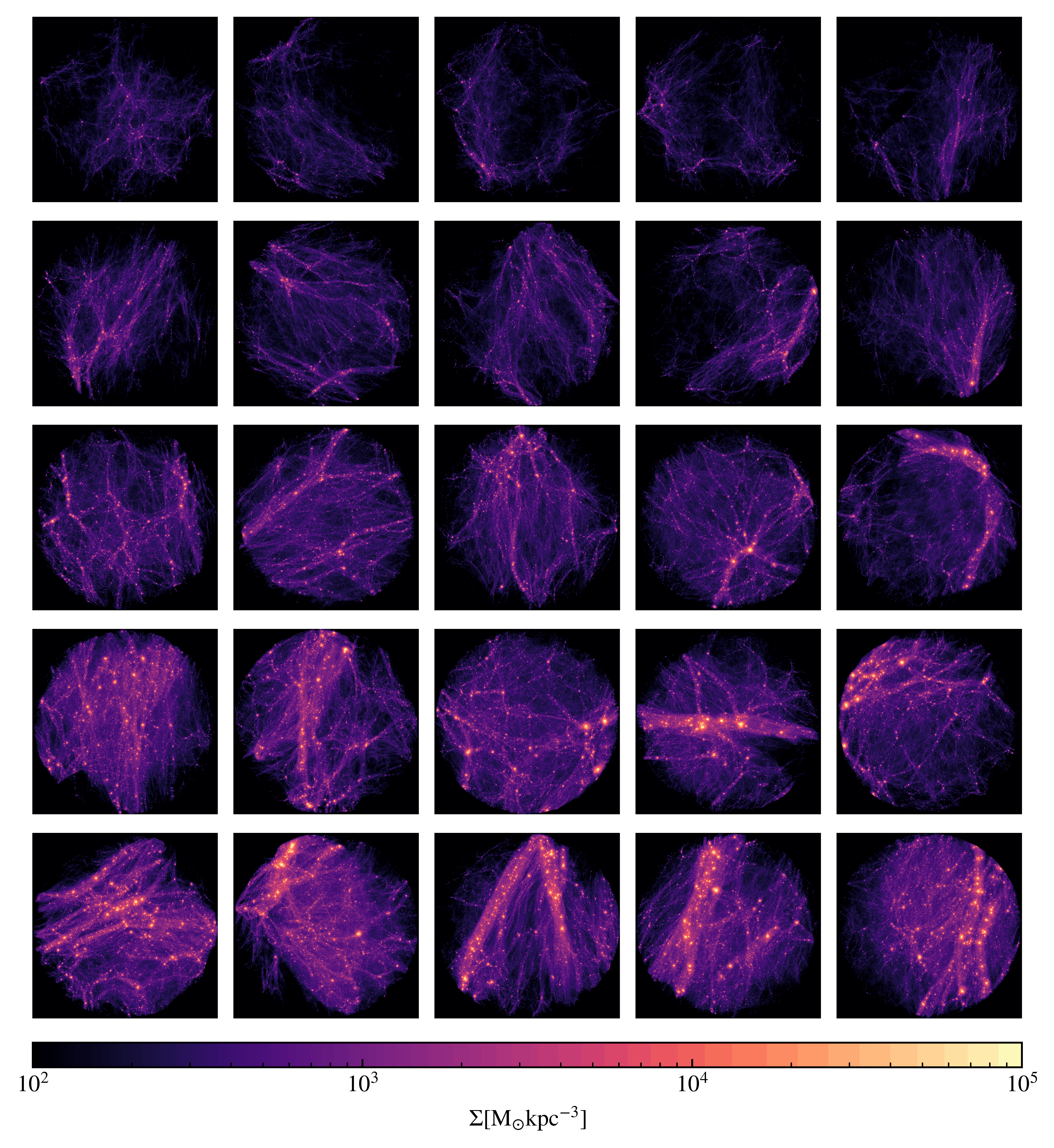}
    \caption{Projections of the total surface mass density in the dark matter only simulations for the 25 selected spherical regions of $r=5\,\cMpc$ at $z=0$. From top to bottom each row corresponds to the $(-2, -1, 0, +1, +2)\sigma$ density band, respectively. From left to right, each column corresponds to the $(-2, -1, 0, +1, +2)\sigma$ value of $N_{\rm LMC}$. For example the bottom row shows the highest density regions that we simulate (the $+2\sigma$ sample), with median density $\log_{10} \dfive = -0.282$ and these regions exhibit, from left to right, $N_{\rm LMC} = (4, 6, 8, 11, 13)$, respectively.}
    \label{fig:region_projections}
\end{figure*}

We select five regions from each of the (-2, -1, 0, +1, +2)$\sigma$ density bands, requiring that for each band the five regions exhibit $N_{\rm LMC}$ corresponding to the (-2, -1, 0, +1, +2)$\sigma$ values of the corresponding $N_{\rm LMC}$ distribution. For any pair of overdensity and $N_{\rm LMC}$, we pick a region at random. Our sample therefore comprises 25 spherical regions of $r=5\,\cMpc$ selected from the $L=400\,\cMpc$ parent volume, and thus follows a total volume comparable to that of a periodic cube of side $L=23.5\,\cMpc$. None of the selected spheres overlap. The centres of the 25 spheres are listed in Table \ref{tab:region_details_appendix}, and their projected total mass density\footnote{For superior clarity we show the surface mass density from the resulting DM-only zoom simulations, rather than the low-resolution parent volume.} is shown in Fig.~\ref{fig:region_projections}, with the rows, from top to bottom, corresponding to (-2, -1, 0, +1, +2)$\sigma$ in density, and the columns, from left to right, corresponding to (-2, -1, 0, +1, +2)$\sigma$ in $N_{\rm LMC}$. The increasing average density from top to bottom is clearly evident.

\subsection{Generation of zoomed initial conditions}
\label{sec:methods:ICs}

We generate multi-resolution `zoomed' initial conditions of the selected regions embedded within the parent volume using the second-order Lagrangian perturbation theory \correction{(2LPT)} software \textsc{ic\_2lpt\_gen} \citep{Jenkins_2010}, again coupled to \textsc{panphasia}. The \textsc{ic\_2lpt\_gen} software assumes that the cosmic matter density is comprised only of baryons and DM, therefore the zoomed ICs are generated with cosmological parameters that differ very slightly from those of the parent box, subsuming the matter density of massive neutrinos into that of the dark energy. The cosmological parameters of the zoomed ICs are therefore: $\Omega_{\rm CDM}=0.256, \Omega_{\rm b}=0.0486, \Omega_{\Lambda}=0.695, h=0.681, \sigma_8=0.807$ and $Y=0.244$. We also make the simplifying assumption that baryons and DM follow the same total matter power spectrum, which is justified as we focus on spatial scales shorter than that on which the baryon power spectrum is suppressed \citep[e.g.][]{VanDaalen_2011}. We generate the linear power spectrum with these parameters using \textsc{class}.

The ICs of each selected region are generated by creating a uniform but multi-resolution particle distribution that represents the unperturbed matter field, to which we apply the (linear + 2LPT) density perturbations of the parent volume supplemented by additional small-scale power down to the appropriate Nyquist frequency. The high-resolution component of each particle distribution has the morphology of the $z=0$ matter distribution within a sphere of $r \simeq 6.5\,\cMpc$ with the same centre as the selected region, when traced back to $z=\infty$. The high-resolution component therefore includes a padding region, corresponding to a spherical annulus of approximate\footnote{The high-resolution component's geometry cannot be specified to arbitrary precision in the unperturbed particle distribution, as it is formed from tiles of particles each in a periodic grid arrangement.} width $r = 1.5\,\cMpc$, that `protects' the selected region from i) contamination by low-resolution composite particles and ii) the influence of the artificial pressure gradient at the boundary of the high-resolution component. The high-resolution component is populated with four DM particles for every gas particle, in a face-replicating grid similar to that adopted by \citet{Richings_2021}. This yields gas and DM particles of approximately equal mass, suppressing the spurious transfer of kinetic energy that can artificially heat galaxy discs \citep{Ludlow_2019}. The remainder of the $L=400\,\cMpc$ cube is populated with collisionless composite particles representing baryons and DM, whose mass increases with distance from the high-resolution region \citep[a detailed explanation of this step is given in Appendix A1.2 of][]{Crain_2009}. The $z=127$ density perturbations are applied by displacing the particles from their unperturbed coordinates using the method described by \citet{Jenkins_2010} and \citet{Jenkins_2013}. We generate ICs for all 25 selected regions at two resolutions, corresponding to high-resolution particle masses of $\sim 10^5\Msun$ (`m5') and $\sim 10^6 \Msun$ (`m6'), to enable convergence testing. The initial H and He mass fractions are assumed to be $X_{\rm H}=0.756$ and $X_{\rm He}=0.244$, respectively. Given the changes between cosmology and IC code, we verified that the zoom ICs faithfully reproduce the structures identified in the parent simulation and preserve the rank ordering of regions by $\delta_5$ and $N_{\rm LMC}$. We also compare the total enclosed mass in $R=5\Mpc$ at $z=0$ between the parent volume and `m5' DM-only zoom simulations and find an average agreement of $0.7\%$ and up to $5\%$ across all regions.

\begin{table*}
	\centering
	\begin{tabular}{lcccccc}
		\hline
		Resolution & $m_{\rm gas}$ & $m_{\rm DM}$ & $\epsilon_{\rm com, gas}$  & $\epsilon_{\rm max, gas}$ & $\epsilon_{\rm com, DM}$  & $\epsilon_{\rm max, DM}$ \\
  	& [M$_{\odot}$] & [M$_{\odot}$] &  [$\ckpc$] & [$\pkpc$] &           [$\ckpc$] & [$\pkpc$] \\
		\hline
		m6 & $1.84 \times 10^{6}$ & $2.42 \times 10^{6}$ & 1.790 & 0.700 & 2.091 & 0.819 \\
		m5 & $2.30 \times 10^{5}$ & $3.03 \times 10^{5}$ & 0.895 & 0.350 & 1.046 & 0.410 \\
		\hline
	\end{tabular}
	\caption{The target masses of gas and DM particles at each of the resolution levels we adopt, with corresponding Plummer-equivalent gravitational softening ($\epsilon_{\rm com}$) length and maximum proper value ($\epsilon_{\rm max}$).}
 	\label{tab:resolution}
\end{table*}

\subsection{Gravity and hydrodynamics}
\label{sec:methods:grav_hydro}

The zoomed ICs are evolved to $z=0$ with \textsc{Swift}. As per the collisionless simulation of the parent volume, gravitational forces were solved using a split fast multipole plus particle-mesh approach. The hydrodynamical equations of motion of the gas were solved using the \textsc{sphenix} \citep{Borrow_2021_SPHENIX} density-energy formulation of smoothed particle hydrodynamics (SPH) adopting the quartic spline kernel with 65 weighted neighbours within the kernel support radius (resolution parameter $\eta = 1.2348$). The SPH solver includes artificial viscosity and conduction terms, with built-in limiters to mitigate numerical energy losses. The time-step limiter of \citet{Durier_2012} is used to ensure accurate evolution of the fluid even during extreme shock events. Time integration uses a leapfrog scheme in which a particle's timestep is the minimum of its gravitational timestep or, if applicable, that dictated by the hydrodynamical Courant-Friedrichs-Lewy condition with parameter $C_{\rm CFL} = 0.2$. The gravitational timestep of a particle with acceleration $a$ is $\Delta t=(0.025\epsilon/a)^{1/2}$, where $\epsilon$ is the gravitational softening length. The timesteps of gas particles cannot exceed that of any of their neighbours by more than a factor of $4$. All high-resolution particles adopt the same  softening length (see Table \ref{tab:resolution}), and the smoothing length is limited to a minimum $h_{\rm min}/\epsilon_{\rm gas} = 10^{-5}$ to avoid artificial runaway collapse in high-density gas \citep{Ploeckinger_2024}.

Stellar and BH particles that form during the simulations do not experience hydrodynamical forces but use an SPH-like kernel to find their neighbours during the operation of subgrid models. BHs use the same $\eta = 1.2348$ kernel as gas particles, whilst stellar particles use a slightly shorter smoothing length ($\eta = 1.1642$). The mass of gas particles can increase due to the donation of mass by neighbouring stellar populations or decrease due to the accretion of mass by neighbouring BHs. Gas particles that exceed their initial mass by a factor of $4$ are split into two particles of equal mass, whilst any reaching half of their initial mass become ineligible for mass accretion onto BHs.

\subsection{Subgrid models}
\label{sec:methods:subgrid}

The simulations appeal to subgrid models to approximate the macroscopic effects of unresolved physical processes. Although we use the same methods as the EAGLE simulations to treat stellar evolution and mass loss, originally developed for the OWLS simulations \citep{Schaye_2010}, the suite of subgrid models used here otherwise represents a substantial advance with respect to the EAGLE model, with several methods in common with the COLIBRE simulations \citep{Schaye_COLIBRE,Chaikin_COLIBRE}. In particular, the methods used for star formation and pre-SN feedback, and the seeding and growth of supermassive BHs, are those developed for COLIBRE, and radiative cooling and heating follows a scheme that allows gas to cool to $T=10\,{\rm K}$. The injection of kinetic and thermal energy from SNe also proceeds similarly to COLIBRE, but differs insofar that a constant fraction of the energy liberated by SNe is coupled to the ISM and a slightly different range of permitted heating temperatures is adopted. Thermal energy from AGN feedback is injected in the same fashion as in COLIBRE, except for the use here of a fixed heating temperature. In contrast to COLIBRE, this model does not include a live dust model, turbulent diffusion, the updated nucleosynthetic yields and SNIa rate prescriptions, nor the jet-mode AGN feedback.

In this sub-section we describe the models used to treat radiative cooling and heating (Section \ref{sec:methods:radiative}), star formation and the ISM (Section \ref{sec:methods:SF_ISM}), the evolution of stellar populations and Type Ia SNe (Section \ref{sec:methods:stellar_evol}), feedback from young stars (Section \ref{sec:methods:esf}) and core collapse SNe (Section \ref{sec:methods:ccsne}), the seeding and growth of BHs (Section \ref{sec:methods:BHs}) and feedback from AGN (Section \ref{sec:methods:AGN}). As many of the adopted methods are common to the EAGLE and COLIBRE projects and are described in detail elsewhere, we present here only concise descriptions, with an emphasis on differences with respect to the COLIBRE model.

\subsubsection{Radiative cooling \& heating}
\label{sec:methods:radiative}

Radiative cooling and heating rates for hydrogen, helium, free electrons from these species and H$_2$ molecules are computed on-the-fly with the non-equilibrium thermochemistry solver \textsc{chimes} \citep{Richings_2014a,Richings_2014b}. The radiative cooling and heating rates of 9 metal species (C, N, O, Ne, Mg, Si, S, Ca and Fe) and dust are pre-computed and tabulated using \textsc{CLOUDY} version 17.03 \citep{cloudy2017}, assuming chemical equilibrium. We largely adopt the treatment of the radiation fields (redshift-dependent UV/X-ray background and interstellar radiation field), cosmic ray rates, and shielding column densities of the fiducial model of \citet{PloeckingerSchaye_2020}, but adopt three changes. First, motivated by the observations of \citet{Lacki2010}, we use a shallower power law scaling of the cosmic ray ionisation rate as a function of the reference column density assumed by the simulations, $\zeta_{\rm CR} \propto N_{\rm ref}^1$, rather than $\zeta_{\rm CR} \propto N_{\rm ref}^{1.4}$. Second, we add a turbulent pressure component (with a 1-dimensional turbulent velocity dispersion of $6~{\rm km\,s}^{-1}$) to the thermal pressure when computing the Jeans column density. Third, we boost the rate of H$_2$ formation on dust grains by a factor that increases from 1 to 100 over the density range $n_{\rm H} = 1-100~{\rm cm}^{-3}$. This is motivated by the simulation's inability to resolve very high density gas clumps in which H$_2$ would form very efficiently. This boost factor for the H$_2$ formation rate is not part of the COLIBRE model, which instead tracks the dust abundance using the live dust model of \citet{Trayford_2025} and enhances H$_2$ via dust growth by accretion. The impact of these changes on the thermal equilibrium temperatures is discussed in detail in \citet{Ploeckinger_2025}. We do not impose an artificial pressure floor, and gas is allowed to cool to a minimum temperature of $10\,{\rm K}$.

\subsubsection{Star formation}
\label{sec:methods:SF_ISM}

We adopt the COLIBRE star formation model detailed by \citet{Nobels_2023}, with the full set of parameters and implementation as detailed in Section 3.3 by \citet{Schaye_COLIBRE}. Gas acquires a non-zero star formation rate (SFR) when it becomes locally unstable against gravitational collapse, specifically when the Jeans mass is smaller than the mass within the SPH kernel. The model has been shown to reproduce the spatially-resolved observed Kennicutt-Schmidt relations for neutral, atomic, and molecular gas, and exhibits good convergence over a wide dynamic range in adopted gas mass resolution. Gas is considered unstable when it satisfies the gravitational instability criterion given by Equation 6 in \citet{Schaye_COLIBRE} which reduces to $\alpha<\alpha_{\rm crit}=1$, where $\alpha$ accounts for thermal and turbulent dispersions. Unstable gas particles are assigned an SFR that follows from the \citet{Schmidt_1959} law implemented via Equation 9 in \citet{Schaye_COLIBRE}, with an efficiency per free-fall time of $\epsilon=0.01$ following observations of star formation within giant molecular clouds \citep[GMCs; e.g.][]{Krumholz_2007}. The SFR is then used as a probability for the stochastic conversion of the gas particle into a stellar particle during each timestep of duration $\Delta t$.

\subsubsection{Stellar evolution and mass loss}
\label{sec:methods:stellar_evol}

We adopt the same \citet{Wiersma_2009b} stellar evolution and mass loss model used by EAGLE and originally developed for OWLS, in which stellar particles are treated as simple stellar populations with a \citet{Chabrier2003} stellar initial mass function (IMF) defined between the limits $0.1$ and $100\,\Msun$. Stellar particles inherit the element abundances of their parent gas particle. We track the time-dependent release of 11 species (H, He, C, N, O, Ne, Mg, Si, S, Ca and Fe) from core collapse and thermonuclear (Type Ia) SNe, and the asymptotic giant branch (AGB) phase of evolved stars. The assumed rate of Type Ia SNe (SNIa) is also very similar to the EAGLE implementation, differing only in the inclusion of an initial delay, $t_{\rm delay}$, between the formation of a stellar particle and the detonation of the first SNIa. The rate of SNIa per unit initial stellar mass formed is therefore:
\begin{equation}
    \dot{N}_{\rm SNIa}(t) = \mathcal{H}(t-t_{\rm delay})\,\frac{v \exp(-2t_{\rm delay}/\tau)}{\tau}\,\exp{\left(-\frac{t-t_{\rm delay}}{\tau}\right)},
    \label{eq: SNIa}
\end{equation}
where $\mathcal{H}$ is the Heaviside step function, $\nu$ is the total number of SNIa per unit initial stellar mass and $\rm{exp}(-t/\tau)/\tau$ is a normalised, empirical delay time distribution function. We use $t_{\rm delay}=40\,\Myr$, $\tau=2\,\Gyr$ and $\nu=1.54\times10^{-3}\,\Msunperyear$. \citet{Schaye_2015} show that these parameters yield a good match to the observed evolution of the cosmic SNIa rate.

\subsubsection{Early stellar feedback}
\label{sec:methods:esf}

The simulations include three pre-supernova feedback processes from massive stars: stellar winds, direct radiation pressure, and H\textsc{ii} regions, using the numerical implementations for COLIBRE which are fully described by \citet[][see also Section 3.6 of \citealt{Schaye_COLIBRE}]{Benitez_Llambay_2025}. These processes are implemented using the Binary Population and Spectral Synthesis (\textsc{bpass}) stellar evolution and spectral synthesis models \citep{Eldridge_2017, StanwayEldridge_2018}, adopting a \citet{Chabrier2003} IMF defined between $0.1$ and $100\,\Msun$. Stellar winds are represented by stochastic momentum transfers to nearby gas, radiation pressure by exerting photon momentum onto the gas based on the local optical depth, and H\textsc{ii} regions by stochastic ionisation and heating of surrounding gas which also temporarily prevents star formation in affected particles.

\subsubsection{Core collapse supernovae}
\label{sec:methods:ccsne}

The core collapse supernovae (CC SNe) feedback model is thermal-kinetic and designed for use in cosmological simulations that \correction{allow gas to cool to $T = 10$ K, partially resolving the cold ISM and hence not needing to impose an effective equation of state. The model injects feedback energy both thermally and kinetically: this} combination of modes generates strong galactic winds and the hot ISM through periodic powerful injections of thermal energy, and turbulence in the ISM through frequent lower energy kinetic kicks \citep{Chaikin_2023}. We broadly follow the approach of COLIBRE \citep[see Section 3.7 of][]{Schaye_COLIBRE} with the exception that we use a fixed feedback energy coupling fraction, $f_{\rm E}=1$, rather than a fraction that is a function of the thermal gas pressure. We elect against tuning the value of the feedback coupling efficiency due to the limitations of the current observed dwarf galaxy samples. The energy released by a stellar particle in a timestep is described by Equation 15 of \citet{Schaye_COLIBRE}, assuming a \citet{Chabrier2003} IMF. A fraction, $f_{\rm kin}=0.1$, of this energy is delivered as stochastic kicks and the remainder as thermal energy. The heating temperature for the thermal feedback is a function of gas density as described by Equation 18 of \citet{Schaye_COLIBRE} for which we adopt a different normalisation density of $n_{\rm H, pivot}=0.25\,\rm cm^{-3}$ and allow the temperature to vary between $10^{6.5}< \Delta T_{\rm SN} < 10^{8}$K.

\subsubsection{Black holes}
\label{sec:methods:BHs}

Our treatment of the seeding and growth of supermassive BHs broadly follows the methods for COLIBRE detailed in Section 3.8 of \citet{Schaye_COLIBRE}, \correction{with the primary difference being that we retain a fixed BH seed mass across resolutions.} The implementation builds on methods introduced by \citet{Springel_BH_2005} and \citet{DiMatteo_2008}. BHs are treated as collisionless particles seeded within haloes that reach a specified mass threshold, and which grow via gas accretion and mergers with other BH particles. We follow the COLIBRE seeding procedure whereby BH particles are seeded in haloes that do not already have a BH particle once their FoF mass exceeds a mass threshold, for Columba this is equal to $M_{\rm FOF, seed} = 10^{10}\,\Msun$. The BH particle is assigned a subgrid mass, $m_{\rm BH}$, used for calculations of BH properties such as its accretion rate, whilst gravitational interactions are computed using the particle's dynamical mass, $m_{\rm BH}^{\rm dyn}$. In Columba, the subgrid mass is initially set to $m_{\rm BH, seed} = 10^{3}\,\Msun$, irrespective of the simulation resolution. The BH mass accretion rate is estimated using the modified form of the Bondi-Hoyle-Lyttleton formula proposed by \citet{Krumholz_2006} for use in turbulent media with vorticity, see Section 3.8.2 of \citet{Schaye_COLIBRE}.

Per \citet[][see also Section 3.8.3 of \citealt{Schaye_COLIBRE}]{Bahe_2022}, BH particle pairs are merged instantly if their separation becomes less than 3 gravitational softening lengths, $\Delta r_{\rm BH} < 3\epsilon_{\rm gas}$, and their relative velocity satisfies $\Delta v_{\rm BH} < 2G(M+m)/\Delta r$, where $M$ and $m$ are the dynamical masses of the two BHs. The less massive of the BHs transfers its properties to the more massive BH particle and is then removed from the simulation.

Real BHs are subject to dynamical friction that drains them of orbital kinetic energy and causes inspiral towards the centre of their host galaxy. Our simulations lack the resolution required to accurately model BH dynamical friction, so we model the net effect of the BH moving towards the local gravitational potential minimum following \citet[][see also Section 3.8.4 of \citealt{Schaye_COLIBRE}]{Bahe_2022}. At each timestep the BH particle is re-positioned to the coordinates of the neighbouring gas particle with the lowest potential. The BH particle is excluded from the calculation of the gravitational potential to avoid becoming trapped by its own potential.

\subsubsection{Active galactic nucleus feedback}
\label{sec:methods:AGN}

We use a single injection mechanism for the AGN feedback energy stemming from gas accretion on BHs that is implemented by thermal heating \citep{BoothSchaye_2009}. This follows the methodology detailed in Section 3.9 of \citet{Schaye_COLIBRE}, with the exception that we adopt a fixed heating temperature rather than scaling it with the BH mass as in COLIBRE. Black holes accumulate feedback energy at a rate proportional to its gas accretion rate \citep[see Equation 34 of][]{Schaye_COLIBRE}, with a coupling efficiency fraction of $\epsilon_{\rm f}=0.05$ in Columba. This energy is stored until it is sufficient to heat at least one neighbouring gas particle by a fixed temperature increment of $\Delta T_{\rm AGN}=10^{9}$K, at which point the energy is released and the BH reservoir is reduced accordingly. This thermal heating scheme prevents excessive numerical radiative losses by ensuring that individual feedback events are sufficiently energetic.

\subsection{Halo \& galaxy identification and sampling}
\label{sec:methods:galaxy_identification}

Haloes are identified in the simulation outputs using the VELOCI\textsc{raptor} structure finder \citep{Elahi_2019}, which first applies a 3-dimensional friends-of-friends (FoF) algorithm to all particles to identify haloes, and then applies a 6-dimensional (position and velocity) FoF algorithm to identify substructures. This approach is advantageous as it enables identification of substructures close to halo centres where the density contrast is small. We use the term ``central galaxies'' (or haloes) to refer to the main substructure of the 3D FoF halo, as identified by VELOCIraptor, while other substructures are considered satellites.  Unless otherwise stated our results only consider central galaxies. The centres of haloes and substructures are defined by the coordinates of their most-bound particle.

The properties of haloes and the galaxies they host are computed using the Spherical Overdensity and Aperture Processor (SOAP) \citep{McGibbon2025}, by aggregating the properties of the particles of which they are comprised, within various spherical apertures about the halo/substructure centres. Some analyses require that haloes are matched with their counterpart from a collisionless (often termed "DM-only") version of the corresponding simulation, which eliminates the influence of baryons on halo growth and structure. This matching is achieved using the bijective particle matching algorithm which has been previously described by \citet{Velliscig_2014}.

\label{sec:concentration_calc}
Analyses in subsequent sections require derived quantities, such as the maximum of the circular velocity profile, $V_{\rm max}$, where the circular velocity profile is $V_{\rm c}(r)=\sqrt{GM(<r)/r}$, or the circular velocity at the virial radius, $V_{200}\equiv V_{\rm c}(R_{200})$. We also consider halo concentration, $c$, which we define as the dimensionless shape parameter of the Navaro-Frenk-White (NFW) density profile \citep{Navarro_1996}, where $c_{\rm NFW}=R_{\rm 200}/R_{\rm s}$ and $R_{\rm s}$ is the NFW scale radius. Rather than performing parametric density profile fits to recover $c_{\rm NFW}$, we estimate the (intrinsic) halo concentration using the ratio of $V_{\rm max}/V_{\rm 200}$ from the matched counterpart haloes in the collisionless simulations, via
\begin{equation}
\frac{V_{\rm max}^{2}}{V_{200}^{2}} \approx 0.216 \frac{\cnfwmath}{A(\cnfwmath)},
\label{eq:c_conversion}
\end{equation}
where, 
\begin{equation}
A(\cnfwmath) \equiv \ln(1+\cnfwmath) - \frac{\cnfwmath}{1 + \cnfwmath}.
\end{equation}
\citet{Bullock_2001} advocate the use of this relationship as it enables recovery of a concentration estimate for low-mass haloes that may lack the particle sampling required for a robust density profile fit. We use the collisionless simulations to eliminate the influence of baryon physics on halo concentrations.

Our analyses consider only those haloes/galaxies that reside within the central $r=5\,\cMpc$ of each region at $z=0$. We have explicitly checked that these regions are uncontaminated by boundary particles, therefore it is guaranteed that the structures we examine are themselves uncontaminated. Aggregating the galaxy populations of all 25 simulations at the present day yields a population of $N(M_\star > 10^7\,\Msun) = 411$ at m6 resolution and $N(M_\star > 10^7\,\Msun) = 448$ at m5 resolution. The m5 resolution simulations yield $N(M_\star > 10^6\,\Msun) = 1321$. In this introductory study we do not construct galaxy samples at earlier epochs, at which time the Lagrangian region of the central $z=0$ sphere does not have a simple morphology, complicating the selection procedure.

%% file: Validation/validation.tex
\section{Model validation and convergence}
\label{sec:validation}

The macroscopic efficiencies of feedback mechanisms are governed by microphysics acting on spatial scales several orders of magnitude below the resolution scale of cosmological simulations \citep[e.g.][]{Orlando_2005}. These efficiencies must therefore be implemented on resolved scales by the adopted subgrid models. The appropriate parametrisation of the subgrid models is difficult to predict a priori, or to infer from observations, and is resolution dependent. It has therefore become common practice to calibrate the parameters of subgrid feedback models, with the aim of having simulations reproduce judiciously-chosen properties of the galaxy population \citep[e.g.][]{Vogelsberger_2014b,Hirschmann_2014,Schaye_2015,Feng_2016,Tremmel_2017,Henden_2018,Pillepich_2018,Dave_2019,Kugel2023}. Such diagnostics commonly include, for example, the galaxy stellar mass function or the inferred stellar mass to halo mass relation, and the central black hole mass - stellar mass relation. 

In the absence of authoritative observational constraints on the galaxy stellar mass function in the low-mass regime \citep[though see][]{Sedgewick_2019}, and similarly for the black hole mass - stellar mass relation, we have not tuned the feedback efficiencies or BH seed mass specifically for these simulations. Instead, we adopt fixed values for the fraction of energy from CC SNe ($f_{\rm E}=1$) and AGN that couples to the ISM ($\epsilon_{\rm f}=0.05$), and the BH seed mass ($m_{\rm BH,seed}=10^3\,\Msun$), as noted in \S\ref{sec:methods}\footnote{Although we do not tune the energy fractions here, had the simulations yielded unrealistic galaxies, we would clearly have adopted a more refined approach.}. Although we did not undertake additional tuning of the parameter choices for these simulations, the underlying subgrid prescriptions inherit parameter choices that were iteratively tested and validated across previous EAGLE \citep{Crain_2015} and COLIBRE \citep{Chaikin_COLIBRE} simulations.

Nevertheless, it is imperative to validate the outcome of the simulations via comparison with observational measurements of fundamental galaxy properties, and to assess how robust these fixed quantities are to changes of numerical resolution. We therefore defer examination of large-scale environmental influences on the dwarf galaxy population to Section \ref{sec:environment}, and first combine the 25 simulations at each resolution to yield a composite galaxy population (albeit one from a collection of regions that are lower density than a random sample, per Section \ref{sec:methods:regions}), in order to examine the galaxy stellar mass function (GSMF, Section \ref{sec:validation:gsmf}), stellar masses as a function of halo mass (Section \ref{sec:validation:smhm}) and galaxy sizes (Section \ref{sec:validation:sizes}).

Our convergence testing is achieved by comparison of the properties of the galaxy population at m5 and m6 resolution using a galaxy formation model with fixed parameters. The comparison thus constitutes a `strong scaling test' in the terminology of \citet[][see their Section 2.2]{Schaye_2015}. Many of the results we examine in Section \ref{sec:environment} follow from the intrinsic properties of DM haloes, which do not depend on the implementation of the galaxy formation model, therefore we examine the numerical convergence of the DM halo mass function and the inner structure of DM haloes separately, in Appendix \ref{appendix:convergence}. Finally, we note that the inclusion of stochastic processes in the adopted subgrid models, such as those governing star formation and feedback, can lead to significant uncertainty in the properties of individual simulated galaxies \citep[e.g.][]{Genel_2019,Keller_2019,DCP_2021,DPC_2022,Borrow_2023}. When studying populations of galaxies, as we do here, this uncertainty is manifested primarily as a source of scatter in scaling relations.

\subsection{The galaxy stellar mass function}
\label{sec:validation:gsmf}

\begin{figure}
    \centering
\includegraphics[width=\columnwidth]{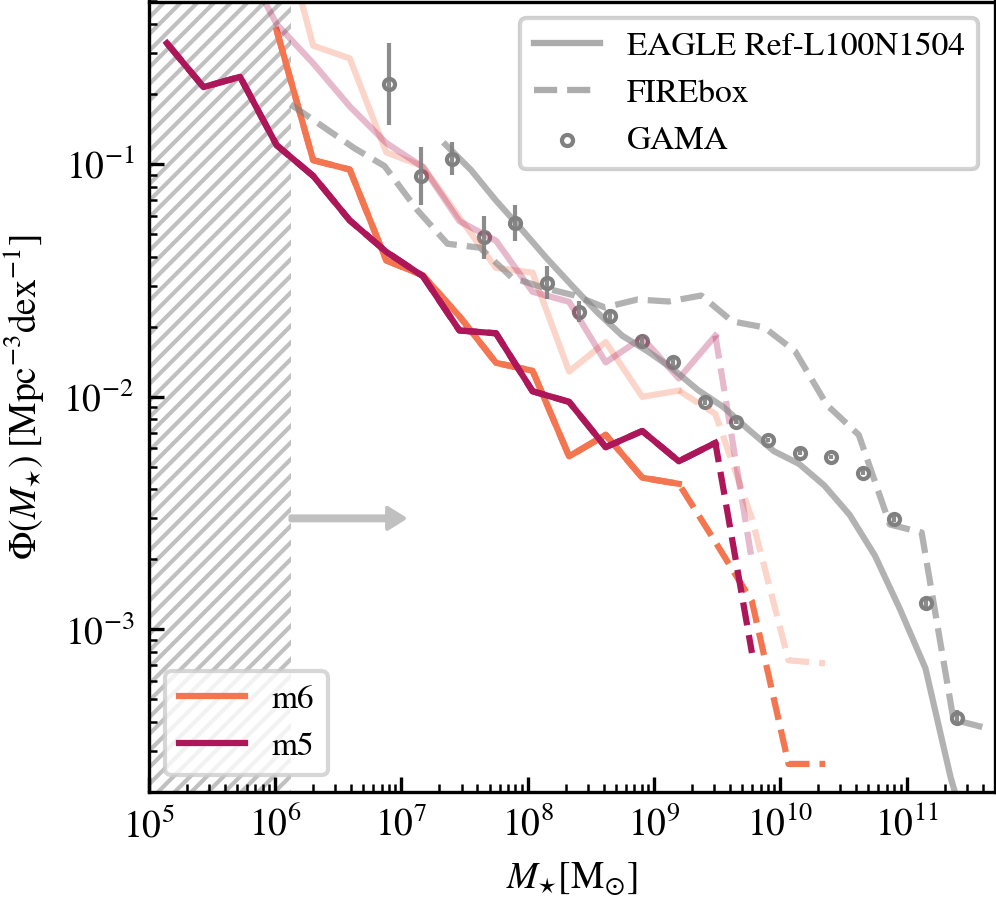}
    \caption{The present-day GSMF of the composite galaxy population from all 25 regions, at m5 (maroon curve) and m6 (orange curve) resolutions, representing a `strong convergence' test. The grey hatched region indicates the stellar mass regime corresponding to 10 or fewer stellar particles (assuming their initial mass) at m5 resolution, the grey arrow extends to the corresponding scale at m6 resolution. The curves are drawn with a dashed line style for bins sampled by fewer than 10 galaxies. Low opacity curves show the GSMFs that result from weighting the number density of galaxies by the reciprocal of their normalised density, $\delta_5$, enabling comparison with GSMFs from representative volumes. Grey symbols show the GSMF and associated error inferred from the GAMA survey \citep{Driver_2022}. Solid and dashed grey curves show the GSMF of the EAGLE \citep{Schaye_2015} and FIREbox \citep{Feldmann_2023} simulations, respectively.
    }
    \label{fig:SMF_convergence}
\end{figure}

Fig. \ref{fig:SMF_convergence} shows the present-day GSMF of the composite galaxy population formed from all 25 spherical regions of $r=5\,\cMpc$. Since we simulate regions that are preferentially low density, one cannot make a direct comparison of the resulting GSMF with observational measurements, nor with GSMFs recovered from simulations of mean-density volumes. We therefore apply a simple correction to the simulated GSMF, rescaling the number density at fixed mass by the reciprocal of the (normalised) mean density of the 25 regions. These rescaled GSMFs are drawn with low-opacity curves, enabling a more meaningful comparison with the GSMF inferred by \citet{Driver_2022} from the fourth data release of the GAMA survey, shown as white circles. For comparison with the GSMF recovered from a simulation of a large periodic volume, we show the present-day GSMF of the EAGLE Ref-L100N1504 simulation \citep{Schaye_2015} as a solid grey curve, and for comparison with a cosmological simulation including an explicit cold ISM phase, we show the FIREbox $z=0$ GSMF \citep{Feldmann_2023} as a dashed grey curve. In both cases the mass function are shown down to a minimum stellar mass corresponding to 10 stellar particles. 

In the regime where galaxies comprise at least 10 or more particles at m6 resolution and the stellar mass bins are sampled by at least 10 galaxies (broadly, $2 \times 10^7\,\Msun \lesssim {\it M}_\star \lesssim 2 \times 10^9\,\Msun$), the simulated GSMF exhibits very good convergence between m5 and m6 resolutions, differing by at most 0.22 dex in stellar mass at fixed number density. This convergence behaviour is superior to that exhibited by the EAGLE model and is likely driven by the inclusion of a cold ISM phase, which stabilises star formation and feedback efficiency across resolutions (see discussion in \citet{Schaye_COLIBRE}). The adoption of the thermal-kinetic CC SNe feedback implementation coupled to the isotropic energy distribution method \citep{Chaikin_2022,Chaikin_2023, Schaye_COLIBRE, Chaikin_COLIBRE} may further contribute to the improved numerical convergence.
At the high-mass end of the well-sampled regime of the GSMF (corresponding to an approximate number density of $5 \times 10^{-3}\,\cMpc^{-3}$) galaxies reach a slightly greater stellar mass in the m5 simulations, likely due to star formation in the progenitors of these galaxies being able to begin at an earlier epoch, and reflecting the fixed AGN model parameters across resolution, which can influence the efficiency of BH growth and feedback regulation.
In the poorly-sampled more massive bins, the handful of galaxies tend to reach greater masses in the lower-resolution m6 simulations. While this may primarily reflect stochastic fluctuations, better convergence in this regime could be achieved by calibrating the BH seed mass as a function of resolution for these simulations. Given the small number of massive galaxies sampled in our simulations, we opted against this additional complexity.

The preferential selection of low density regions leads to the unscaled GSMFs residing at a lower number density at fixed stellar mass, relative to the GAMA GSMF. In the regime where simulated galaxies are well resolved and the stellar mass bins well sampled, the rescaled GSMFs of the simulations reproduce the GAMA GSMF well, with the stellar mass at fixed number density differing by $\lesssim 0.22\,{\rm dex}$. This is comparable to the agreement achieved by the EAGLE flagship Ref-L100N1504 simulation, but here this level of agreement extends to stellar masses a factor of 8 lower for the m5 simulations. Even after rescaling, the number density of galaxies of mass $M_\star \gtrsim 3 \times 10^9\,\Msun$ is significantly lower than inferred by GAMA: this stems from a paucity of massive DM haloes in the selected regions by construction, as can be seen by comparison of the halo mass functions of the regions with that of mean density volumes (see Fig.~\ref{fig:HMF_convergence}).

\subsection{Stellar mass as a function of halo mass}
\label{sec:validation:smhm}

\begin{figure}
    \centering
	\includegraphics[width=\columnwidth]{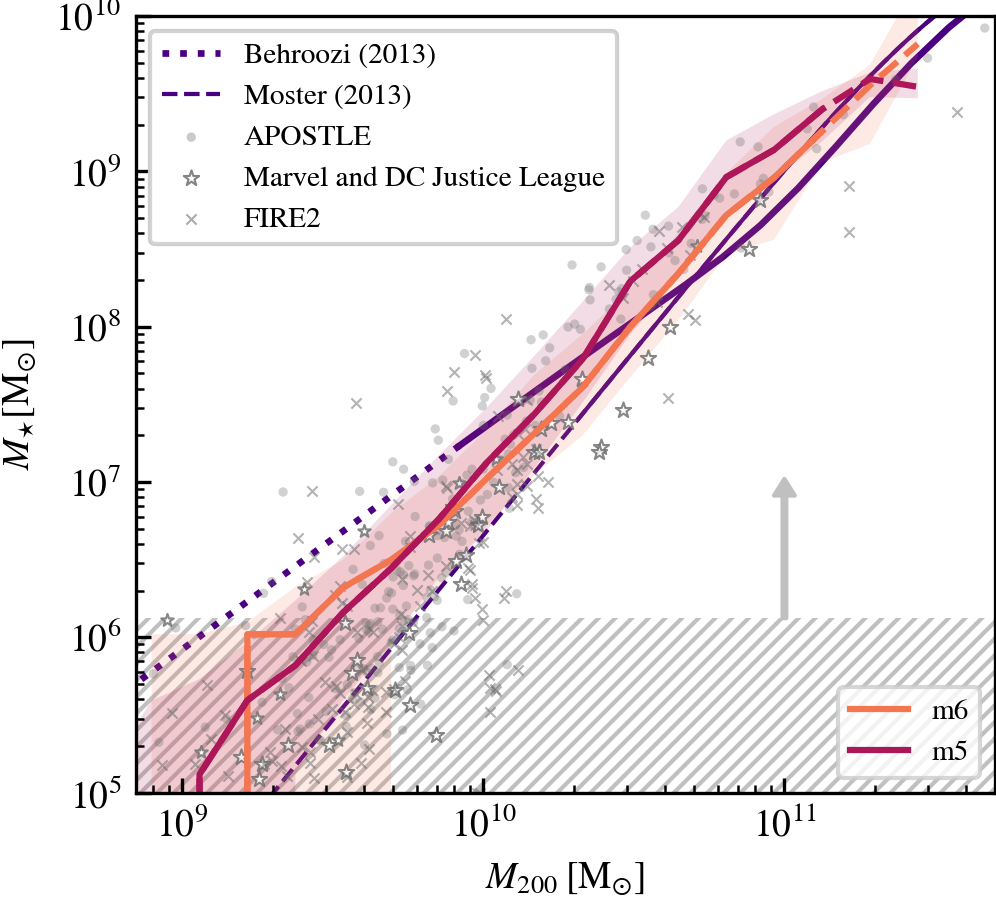}
    \caption{The present-day stellar mass - halo mass scaling relation of central galaxies for the composite population of all 25 regions, at m5 (maroon curve) and m6 (orange curve) resolutions, representing a `strong convergence' test. Line styles and hatching follow the conventions introduced in Fig.~\ref{fig:SMF_convergence}. Shaded regions denote the $10^{\rm th} - 90^{\rm th}$ percentile scatter in $M_\star$ at fixed $M_{200}$. Indigo curves show the SMHM relations of the semi-empirical models of \citet{Behroozi_2013} and \citet{Moster_2013}, which transition to solid lines in the mass range to which they are fit. Individual central galaxies from various models are shown by the grey points, including APOSTLE \citep[filled circles,][]{Sawala_2016}, Marvel and DC Justice League \citep[stars,][]{Applebaum_2021}, and FIRE2 zoom-in simulations \citep[crosses,][]{Wheeler_2015, Fitts_2017, Hopkins_2018, Wheeler_2019}.    }
    \label{fig:SHMR_convergence}
\end{figure}

To first order, the abundance and growth of dark matter haloes are simply a function of the assumed cosmological parameters. The stellar mass of the galaxies that forms within them is, however, very sensitive to the baryonic physics included in the simulation. The stellar mass - halo mass (SMHM) scaling relation has emerged as a popular diagnostic, in part because it can be considered as a halo mass-dependent galaxy formation `efficiency' on the scale of individual haloes. We include the SMHM here as a validation diagnostic as the GSMF can be thought of as a convolution of the DM HMF with the SMHM relation. Moreover, most numerical simulations in the low-mass dwarf galaxy regime maximise spatial resolution by adopting zoomed initial conditions that focus on the environments of individual galaxies, so the SMHM relation is a more useful diagnostic than the GSMF when comparing our simulations to others.

Fig. \ref{fig:SHMR_convergence} shows the present-day SMHM of the composite population of field galaxies formed from all 25 simulated regions. Satellite galaxies, whose evolution is potentially affected by ram pressure stripping of gas and mass loss due to tidal stripping, are excluded\footnote{Due to the relatively low mass of our field sample and the resolution limit of the simulations, the fraction of satellites is small; only 11\% of all galaxies with $M_{\star}>10^5 \Msun$ are satellites.} in order to illustrate the scatter about the relation due only to processes `internal' to the galaxy's parent DM halo, however we note that we have not attempted to identify and exclude `backsplash galaxies', i.e. present-day central galaxies that were satellite galaxies at an earlier time. We do not expect a significant population of such galaxies, because there are no massive haloes in our simulated regions. We show the median stellar mass as a function of halo mass in bins of width 0.2 dex, and the corresponding $10^{\rm th} - 90^{\rm th}$ percentile scatter.

The indigo curves denote the present-day SMHM relations 
from the abundance matching models of \citet{Behroozi_2013}\footnote{We convert halo masses in this work, which were computed based on the virial overdensity criterion of \citet{BryanNorman1998}, to $\mhalo$ assuming $\mhalo=M_{\rm vir}/1.2$} and \citet{Moster_2013}. These models are essentially unconstrained for $M_\star \lesssim 10^7\,\Msun$, and in this regime we simply extrapolate the SMHM relation from the resolved regime, drawing the corresponding curves with dotted and dashed line styles, respectively. The uncertainty on the SMHM relation in this regime is a hindering factor in testing predictions for the properties of dwarf galaxies from numerical simulations. \citet{Sales_2022} collate predictions for the SMHM relation from various cosmological hydrodynamical simulations of dwarf galaxies, revealing similar stellar masses at fixed halo mass (for $\mhalo \gtrsim 10^9\,\Msun$, albeit with large scatter) and broad compatibility with extrapolations of the predictions of semi-empirical models. However, the slope and scatter of the SMHM relation recovered from the simulations considered by \citet{Sales_2022} exhibit significant variance. We therefore plot with grey symbols individual central galaxies from high-resolution simulations: filled circles are from the APOSTLE suite \citep{Sawala_2016} of zoomed simulations of Local Group (LG)-like environments; open stars are from the Marvel and DC Justice League suites \citep{Applebaum_2021} that simulate dwarfs in cosmic sheets and MW-like environments, respectively; crosses denote the Latte \citep{Wetzel_2016} and ELVIS \citep{Garrison-Kimmel_2014} zoom simulations of MW and LG-like environments, in addition to zoom simulations of isolated dwarf galaxies, all evolved with the FIRE-2 model \citep{Wheeler_2015,Fitts_2017, Hopkins_2018, Wheeler_2019}. 

The good convergence of the GSMF between the m5 and m6 resolutions evident in Fig.~\ref{fig:SMF_convergence} translates into similarly good convergence of the SMHM relation. In the regime for which galaxies are resolved by at least 10 stellar particles at m6 resolution ($M_\star \gtrsim 2 \times 10^7\,\Msun$) and halo mass bins are sampled by at least 10 galaxies ($M_{200} \lesssim 10^{11}\,\Msun$), the scaling relation differs in stellar mass by at most 0.29 dex at fixed halo mass. As one might infer from the analysis of the GSMF, the poorest correspondence is for the relatively massive galaxies, with the m5 simulations exhibiting greater stellar masses at fixed halo mass, until one reaches a mass scale of $M_{200} \sim 10^{11}\,\Msun$. Above this mass, the central BHs in these galaxies tend to have greater characteristic masses at m5 resolution, which may contribute to stronger regulation of stellar mass relative to the m6 simulations. However, given the small number of objects in this regime, this should be interpreted with caution. In the low mass regime ($M_{200} \lesssim 10^{10}\,\Msun$), simulations of low-mass dwarf galaxies exhibit significant scatter in stellar mass at fixed halo mass, arguably in part owing to the absence of authoritative observational constraints to which the models can be calibrated. The SMHM relations produced by our simulations roughly bisect the extrapolated semi-empirical relations of \citet{Moster_2013} and \citet{Behroozi_2013}, and lie within the scatter of the relations traced by the APOSTLE, Marvel, DC Justice League and FIRE-2 simulations. 

\subsection{Galaxy sizes}
\label{sec:validation:sizes}

\begin{figure}
    \centering
	\includegraphics[width=\columnwidth]{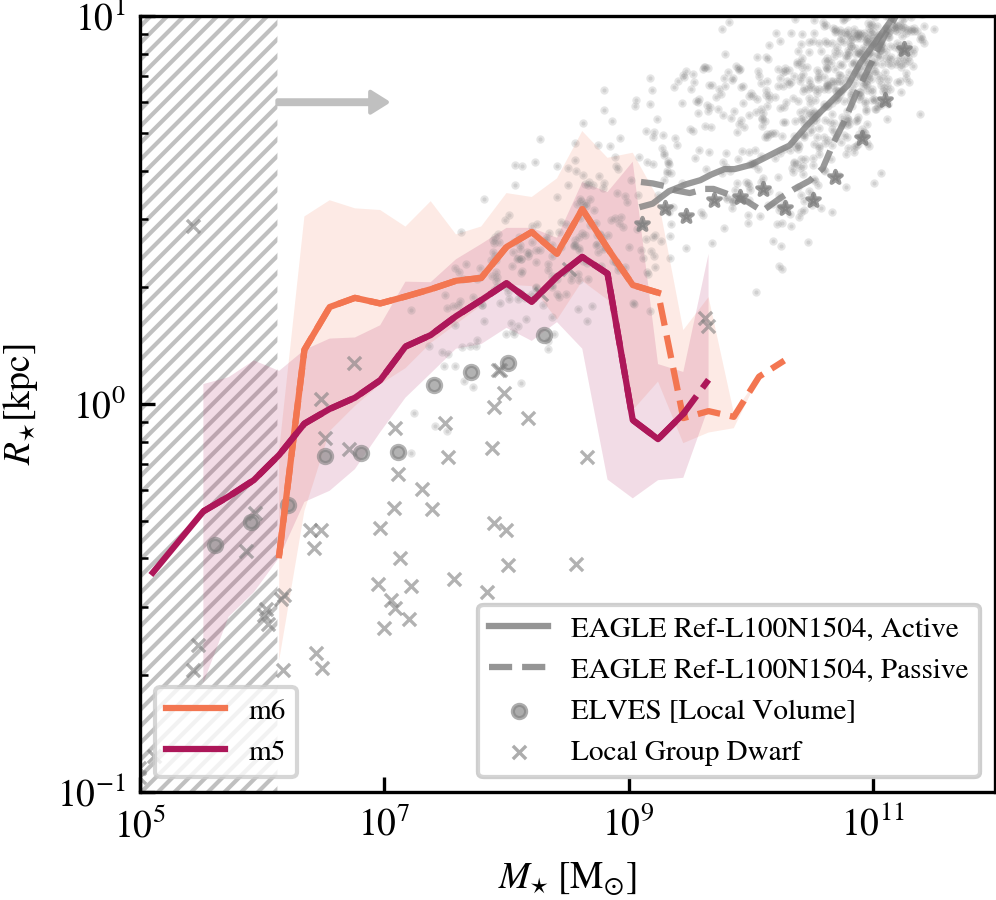}
    \caption{The present-day galaxy size (3D half-mass radii) - stellar mass scaling relation of the composite galaxy population from all 25 regions, at m5 (maroon curve) and m6 (orange curve) resolutions, providing a `strong scaling' test. Shaded regions denote the $10^{\rm th} - 90^{\rm th}$ percentile scatter in $R_\star$ at fixed $M_\star$. Line styles and hatching follow the conventions introduced in Fig.~\ref{fig:SMF_convergence}.
    The median sizes of galaxies sampled by 100 or more particles in the EAGLE simulation are shown with grey curves \citep{Furlong_2017}. Grey points show observational measurements of individual galaxies from \citet{Trujillo_2020}. Grey star symbols show median measurements from \citet{Hardwick_2022}. Circles show medians of the ELVES Local Volume dwarf galaxies from \citet{Carlsten_2021}. Crosses show Local Group dwarf galaxies from \citet{Pace_2024}. Projected observational sizes have been converted to 3D sizes using  $R_{\rm 3D}=\frac{4}{3}R_{\rm 2D}$.}
    \label{fig:sizes_convergence}
\end{figure}

\citet{Crain_2015} showed that it was possible to reproduce accurately the present-day GSMF in a version of the EAGLE simulations adopting an uncalibrated feedback coupling efficiency (i.e. $f_{\rm E} = 1$), as we adopt here, but at the cost of a poor reproduction of the scaling relation connecting the size and stellar mass of galaxies. This shortcoming stemmed primarily from spurious radiative losses in high-density gas, resulting in the formation of unrealistically compact galaxy nuclei. We therefore consider galaxy sizes a critical validation diagnostic, and define size as the 3-dimensional stellar half-mass radius, $R_\star$. Fig.~\ref{fig:sizes_convergence} shows the median size of all present-day galaxies in our composite sample as a function of stellar mass in bins of width 0.2 dex, and the corresponding $10^{\rm th} - 90^{\rm th}$ percentile scatter. The solid and dashed grey curves show the corresponding scaling relations for active and passive galaxies, respectively, with mass $M_\star \gtrsim 10^9\,\Msun$ taken from the EAGLE Ref-L100N1504 simulation, as presented by \citet{Furlong_2017}. We also present observed galaxy sizes for reference, and use the relation $R_{\rm 3D}=\frac{4}{3}R_{\rm 2D}$ to convert between observed 2D (projected) and 3D sizes \citep{Wolf2010}. Faint grey dots denote measurements of projected galaxy sizes from the IAC Stripe 82 Legacy Project by \citet{Trujillo_2020}, grey stars correspond to size measurements of a sub-sample of xGASS galaxies with robust rotational velocity measurements \citep{Hardwick_2022}, grey circles show local volume galaxies from the ELVES survey \citep{Carlsten_2021}, and grey crosses Local Group field galaxies \citep{Pace_2024} \footnote{This study used the Local Volume Database, \url{https://github.com/apace7/local_volume_database }.}.

In the mass regime for which galaxies are well resolved in the m5 and m6 realisations, the simulations yield a size-mass relation that is broadly monotonic up to a mass scale of $M_\star \simeq 5 \times 10^8\,\Msun$. In m5, the change in size across two orders of magnitude in mass is comparable to that seen in observational samples, although the simulated galaxies are systematically larger at fixed mass. The visual impression of a flatter relation in the simulations primarily reflects this vertical offset rather than a substantially different slope. Beyond this scale, at both resolutions galaxies quickly become compact, qualitatively similar to the unrealistically compact sizes reported by \citet{Crain_2015} for variants of the EAGLE model that do not scale $f_{\rm E}$ with the density of natal gas, leading to numerically-inefficient feedback due to artificial radiative losses. This likely signals that additional energy input ($f_{\rm E} > 1$) is required in the progenitors of these more massive present-day galaxies to offset artificial losses. The sizes of the most massive galaxies ($M_\star \gtrsim 10^9\,\Msun$ at m5 and $M_\star \gtrsim 3 \times 10^9\,\Msun$ at m6) increase mildly as a function of stellar mass, possibly indicating that artificial radiative losses from SN feedback are (at least partially) offset by the onset of effective AGN feedback.

The m6 simulations yield larger galaxies at fixed mass, a result qualitatively consistent with the strong convergence test of galaxy sizes in EAGLE reported by \citet[][see their appendix B]{Furlong_2017}. At the mass scale corresponding to 10 particles at m6 resolution, $M_\star \simeq 2 \times 10^7\,\Msun$, the median galaxy size at m6 resolution ($1.92\,\kpc$) is approximately 1.5 times that at m5 ($1.39\,\kpc$). The offset declines towards greater masses: at $M_\star = 3 \times 10^8\,\Msun$ the corresponding sizes are $2.40\,\kpc$ at m6 and $2.22\,\kpc$ at m5. This highlights that galaxy sizes require significantly better particle sampling to achieve convergence than stellar masses \citep[see also e.g.][]{Furlong_2017,Pillepich_2018b,Ludlow_2019,Ludlow_2020}, and that they are more sensitive to the details of the galaxy formation model than galaxy (stellar) masses.

The simulated galaxies are systematically larger than inferred by \citet{Carlsten_2021} from observations of dwarf galaxies in the Local Group. It is not trivial to compare 3-dimensional half-mass radii with isophotal sizes \citep[see e.g.][]{Genel_2018, DeGraaff_2022}, and although we convert the observed projected radii into 3D, this conversion remains approximate. Even so, a na\"ive comparison indicates that in the mass range $2 \times 10^6\,\Msun < \mathit{M}_\star < 5 \times 10^8\,\Msun$, galaxies in the m5 simulations are up to twice the size of observed counterparts from \citet{Carlsten_2021} and up to a factor of 3.5 times larger than the binned median of the \citet{Pace_2024} sample. This indicates that the simulations do not reproduce the smaller observed galaxy sizes; even the $10^{\rm th}$ percentile simulated sizes lie well above the more compact Local Group galaxies in the observational sample.

Although the sizes of galaxies with $M_\star \gtrsim 3 \times 10^8\,\Msun$ are clearly inconsistent with observational measurements, warranting care when interpreting related properties such as their star formation histories, the simulations reproduce a size-mass relation with a similar slope to observations at lower masses. At fixed stellar mass, the offset in size between the m5 simulations and the \citet{Carlsten_2021} measurements is comparable to, or smaller than, the difference between the \citet{Trujillo_2020} and \citet{Carlsten_2021} measurements (in the mass regime for which these samples overlap).
The simulations therefore differ from the \citet{Carlsten_2021} measurements by less than what might reasonably be inferred as the systematic uncertainty on the latter. We also caution that incompleteness of the data can play a role here as more extended, diffuse dwarf galaxies are harder to discover; e.g. ultra diffuse Antlia II around the Milky Way was missing up until recent years and was discovered largely due to \textit{Gaia} proper motion measurements \citep{Torrealba2019}. Moreover, the data in the lower mass range is from dwarf galaxies around higher mass hosts which is a different environment from our sample of simulated dwarf galaxies. Overall, we can conclude that while the simulations reproduce the qualitative trend of increasing size with stellar mass below $M_\star \lesssim 5 \times 10^8\,\Msun$, they likely remain systematically too large, particularly relative to Local Group field dwarfs.

%% file: Results/results.tex
\section{The influence of large-scale environment on isolated dwarf galaxies}
\label{sec:environment}

In this section we turn to one of the main questions motivating the simulation suite: what is the influence of cosmic environment on the properties of dwarf galaxies and their parent DM haloes? In the light of this question, we examine fundamental properties of our simulated dwarf galaxy population as a function of their environment defined by the mass density within $r=5\,\cMpc$ spheres, $\dfive$. As illustrated by Fig.~\ref{fig:region_projections}, the five bands of $\dfive$ (denoted by rows in the figure) broadly correspond to differing characteristic cosmic environments, from void-like regions to filamentary structures. Unless otherwise stated, we focus exclusively on the simulations at m5 resolution in this section. 

\subsection{The halo mass function}
\label{sec:environment:hmf}

\begin{figure}
    \centering
    \includegraphics{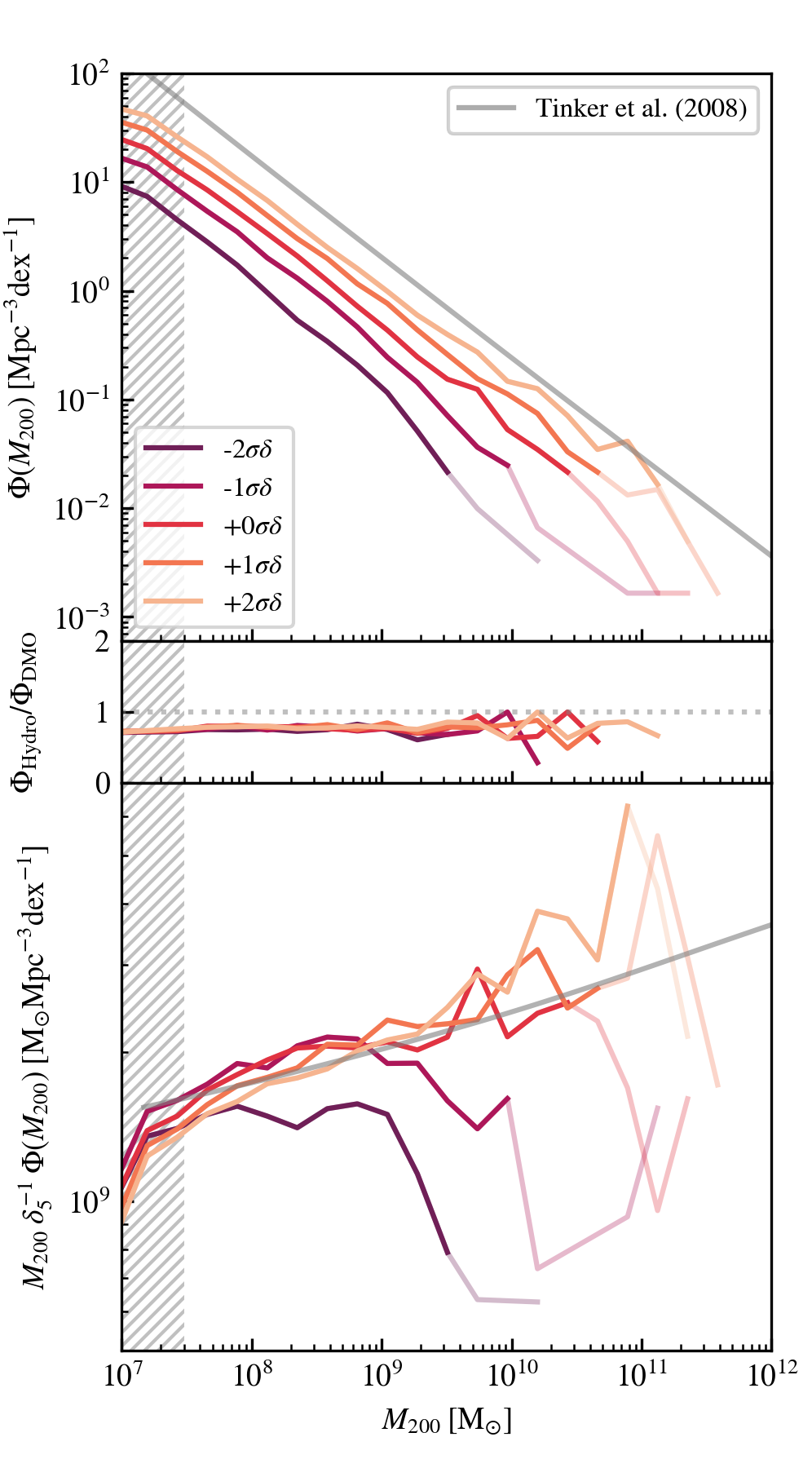}
    \caption{\textit{Upper panel}: The aggregated present-day halo mass functions of the five regions comprising each of the five density bands ($-2\sigma$, ..., $+2\sigma$) at m5 resolution, with each band denoted by a coloured curve. Only field (central) objects are considered. Grey hatching indicates the halo mass regime corresponding to fewer than 100 DM particles at m5 resolution. Curves are drawn with a reduced opacity for bins sampled by fewer than 10 haloes. The grey curve shows the halo mass function of \citet{Tinker_2008} derived from purely collisionless simulations. Higher density regions exhibit a systematically greater number density of haloes at fixed halo mass. 
    \textit{Centre panel}: The ratio of the halo number density at fixed halo mass, $\Phi(\mhalo)$, realised in hydrodynamical simulations, to that in their purely collisionless counterpart simulations (DMO). The curve is drawn for bins sampled by at least 10 haloes. The ratio illustrates the suppression of the HMF due to baryonic processes, which systematically reduce the mass of haloes. \textit{Lower panel}: As per the upper panel, but the HMFs have been multiplied by $\mhalo$ to reduce the dynamic range of the $y$-axis, and normalised by $\delta_5$ to remove the first-order influence of enclosed density on halo number density, highlighting that the offset from the \citet{Tinker_2008} HMF visible in the upper panel is primarily a consequence of each region's underdensity. The most massive halo formed in each region is a strong function of $\delta_5$.}
    \label{fig:HMF}
\end{figure}

A fundamental prediction of CDM cosmogonies is that the rate at which a region's halo population evolves is dependent on its density \citep[e.g.][]{Frenk_1988,Sheth_Tormen_2002}. We therefore show in the upper panel of Fig.~\ref{fig:HMF} the present-day halo mass functions (HMFs) of our hydrodynamical simulations as a function of  $\dfive$. We amalgamate the halo populations of the five regions comprising each band of $\dfive$ (i.e. $-2\sigma$, ..., $+2\sigma$), normalising by their combined volume. The grey curve is a universal mean density HMF fit for the adopted cosmogony of our simulations, created with the \textsc{colossus} package \citep{Diemer_2018} based on the \citet{Tinker_2008} HMFs assuming spherical overdensity (SO) haloes with mean internal density contrast $\Delta_{\rm crit}=200$. Note that their halo centre definition (density peak) differs slightly from the most bound particle centre used in our simulations. This HMF is based on fits to ensembles of collisionless simulations, and does not account for the influence of galaxy formation on the mass and internal structure of haloes; such effects are halo mass dependent, and are the subject of intense study \citep[see e.g.][]{Stanek_2009,Duffy_2010,Velliscig_2014,Schaller_2015,Bocquet_2016,Lovell_et_al_2018}. 

As expected, given that the enclosed densities of all simulated regions are less than the cosmic mean density (i.e. $\delta_5 < 1$), the HMFs of all five density bands are shifted to lower number densities at fixed $M_{200}$ than the universal HMFs, with the offset being greatest for the lowest-density band. In the low-mass regime, for which the HMF is well approximated by a power law, the offset in number density at fixed $M_{200}$ is a monotonic function of enclosed density: at $M_{200}=10^8\,\Msun$, the bands are offset by (1.15, 0.81, 0.61, 0.44, 0.31)$\,{\rm dex}$, respectively, for (-2, -1, 0, +1, +2)$\sigma$ regions, from the \citet{Tinker_2008} HMF. The influence of the density of the regions is more readily seen in the lower panel, in which the HMFs of the regions comprising each band have been divided by the enclosed density, $\dfive$, and multiplied by $M_{200}$ to reduce the y-axis dynamic range, thus better revealing differences in the `specific' halo number density of the density bands. With the influence of density `normalised out', the low-mass power-law regime of the HMFs ($M_{200} < 10^9\,\Msun$) more closely resembles the universal HMF. There remains some scatter about the universal HMF, with a weak systematic trend apparent such that the higher-density regions tend to exhibit a slightly lower specific halo number density than lower-density regions \citep[as also found in the GIMIC simulations, see][]{Crain_2009}. However, the lowest-density region retains a low specific halo number density after normalisation, and fully characterising whether there is a genuine systematic trend in specific halo number density as a function of overdensity would require examination of a larger sample of regions. The slope of the low-mass end of the HMF is also broadly insensitive to the density of the region. \citet{Fattahi_2020} report similar results, finding a correlation with little scatter between the number of low mass haloes and the total mass within a sphere of $r=3\,\cMpc$ centred on the Local Group-like environments. The lower panel of Fig.~\ref{fig:HMF} also highlights that the characteristic mass above which the HMF ceases to be well described by a power law is broadly a function of $\dfive$: more massive haloes in our simulations are only found in the higher density regions, consistent with prior analytical and numerical studies of halo growth \citep[e.g.][]{Mo_1996,Sheth_Tormen_2002}. 

The scale of the influence of baryonic processes on the HMF is illustrated by the centre panel, which shows the ratio of the halo number density (at fixed halo mass) of the hydrodynamical simulations to that derived from their counterpart simulations with purely collisionless dynamics. In the halo mass regime for which the HMF is well described by a power law, the ratio of the halo number density is approximately $\Phi_{\rm Hydro}/\Phi_{\rm DMO} \simeq 0.8$, where $\Phi(\mhalo) \equiv \mathrm{d}n(\mhalo)/\mathrm{d}\log_{10}(\mhalo)$. Physically, this effect should more appropriately be considered as a shift in halo mass at fixed number density, i.e. a horizontal rather than vertical shift of the curves. \citet{Sawala_2013} showed that the reduction of the mass of haloes below $\sim10^{12}\Msun$ when baryon physics is included is primarily a consequence of the unbinding of circumgalactic gas due to heating by the metagalactic UV/X-ray background radiation that permeates the cosmos after the epoch of reionisation, and the subsequent reduction in the halo mass accretion rates.

\subsection{The galaxy stellar mass function}

\begin{figure}
    \centering
	\includegraphics{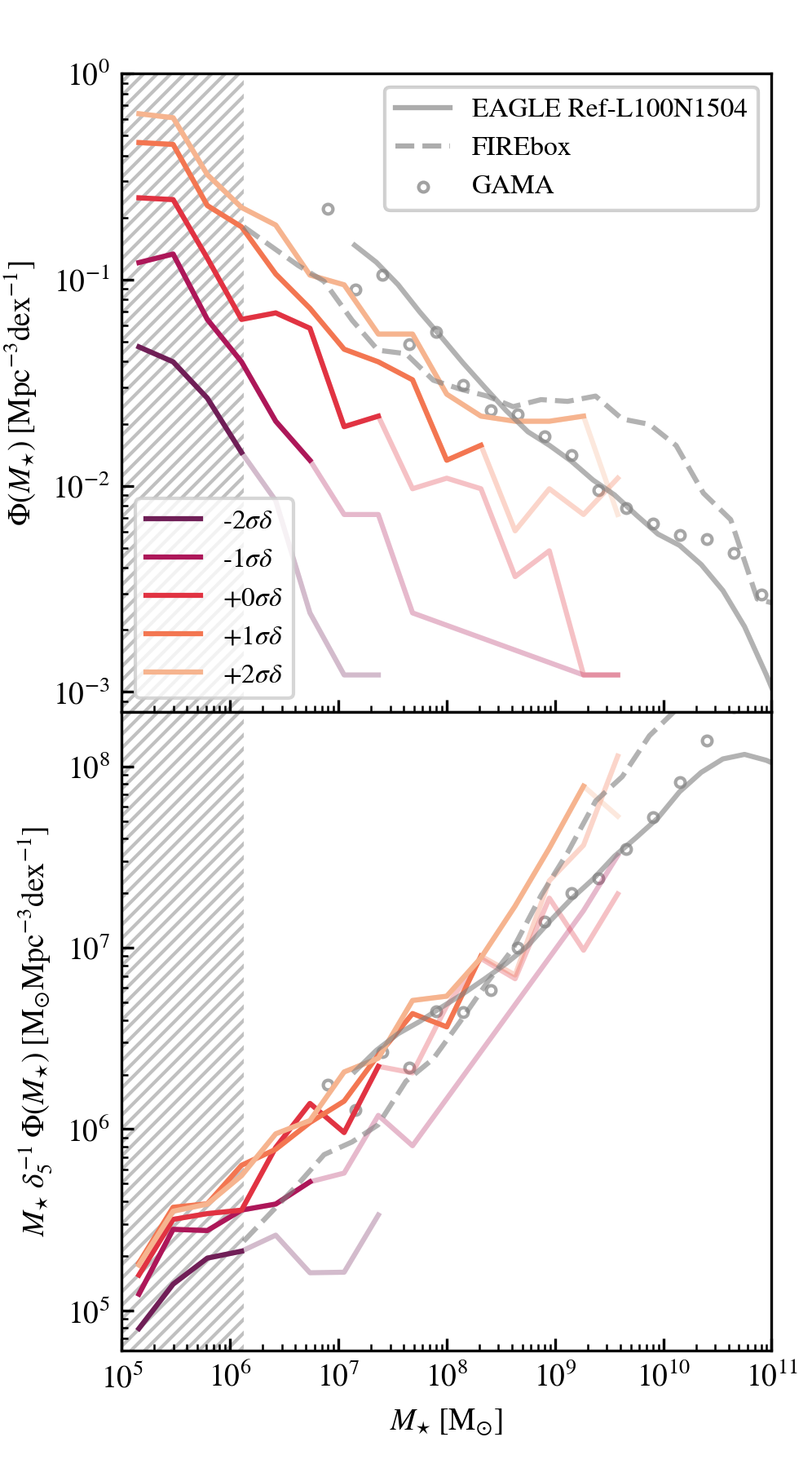}
    \caption{\textit{Upper panel}: The aggregated present-day central galaxy stellar mass functions of the five regions comprising each of the five density bands, with each band denoted by a coloured curve. Grey hatching indicates the stellar mass regime corresponding to fewer than 10 stellar particles at m5 resolution. Curves are drawn with a reduced opacity for bins sampled by fewer than 10 galaxies. Grey symbols show the GSMF inferred from the GAMA survey \citep{Driver_2022}. Dotted and dashed grey curves show the GSMF of the EAGLE \citep{Schaye_2015} and FIREbox \citep{Feldmann_2023} simulations, respectively. In analogy with the HMF, higher density regions exhibit a systematically greater number density of galaxies at fixed stellar mass. 
    \textit{Lower panel}: As per the upper panel, but as in Fig.~\ref{fig:HMF}, the GSMFs have been multiplied by $\mstar$, and normalised by $\delta_5$ to remove the effect of enclosed density on halo number density. As with the HMF (Fig.~\ref{fig:HMF}) the offset from the GAMA GSMF visible in the upper panel follows primarily from the deviation of each region's mass density from the cosmic average.
    }
    \label{fig:GSMF}
\end{figure}

The upper panel of Fig.~\ref{fig:GSMF} shows the present-day galaxy stellar mass functions (GSMFs) at m5 resolution. As in Fig.~\ref{fig:HMF}, each coloured curve denotes the GSMF of a different density band, $\dfive$, generated by aggregating the galaxy populations of the band's five constituent simulations. The solid and dashed grey curves show, respectively, the present-day GSMFs of the EAGLE Ref-L100N1504 \citep{Schaye_2015} and the FIREbox \citep{Feldmann_2023} simulations. Open grey symbols denote the GSMF recovered from the fourth data release of the GAMA survey \citep{Driver_2022}.

The systematic offset of the HMFs as a function of $\dfive$ shown in Fig.~\ref{fig:HMF} translates into a similar offset of the GSMFs of the regions comprising the five density bands, as the formation of fewer haloes in low-density environments results in the formation of fewer galaxies. Just as the HMFs of the simulations are offset from the mean density HMFs, their GSMFs are also in general offset to lower characteristic number densities than those of the simulations of mean density volumes, and that from GAMA. We again examine the effect of environment on the GSMF more closely by showing in the lower panel of Fig.~\ref{fig:GSMF} the GSMFs multiplied by the bin mass (in this case, $M_\star$) and normalised by $\dfive$. This panel illustrates that the slope of the GSMF in the low-mass regime is broadly insensitive to the environment defined by $\dfive$; in common with the HMF, by `normalising out' the enclosed density, the GSMFs of each ensemble of simulations comprising the five density bands more closely resembles the GAMA GSMF, indicating that, to first order, the low normalisation of the low-mass regime of the GSMF in low-density regions is a consequence of that lower density. There is however a small residual offset, such that the density-normalised GSMFs of the lowest-density bands ($-2\sigma$ and $-1\sigma$) remain offset from those of the $0\sigma$, $+1\sigma$ and $+2\sigma$ bands. The stellar mass formed per unit total mass is therefore slightly but significantly lower in the lowest density regions. This follows primarily from the analogous differences in halo abundances at the corresponding halo mass scale, seen in the lower panel of Fig.~\ref{fig:HMF}: galaxies of $M_{*} \simeq 10^6 \Msun$ are hosted by haloes of mass $\mhalo \gtrsim 10^9 \Msun$, the scale at which clear systematic differences between the density bands become apparent in the density-normalised HMF. 

\subsection{Stellar mass as a function of halo mass}

\begin{figure}
    \centering
    \includegraphics[width=\columnwidth]{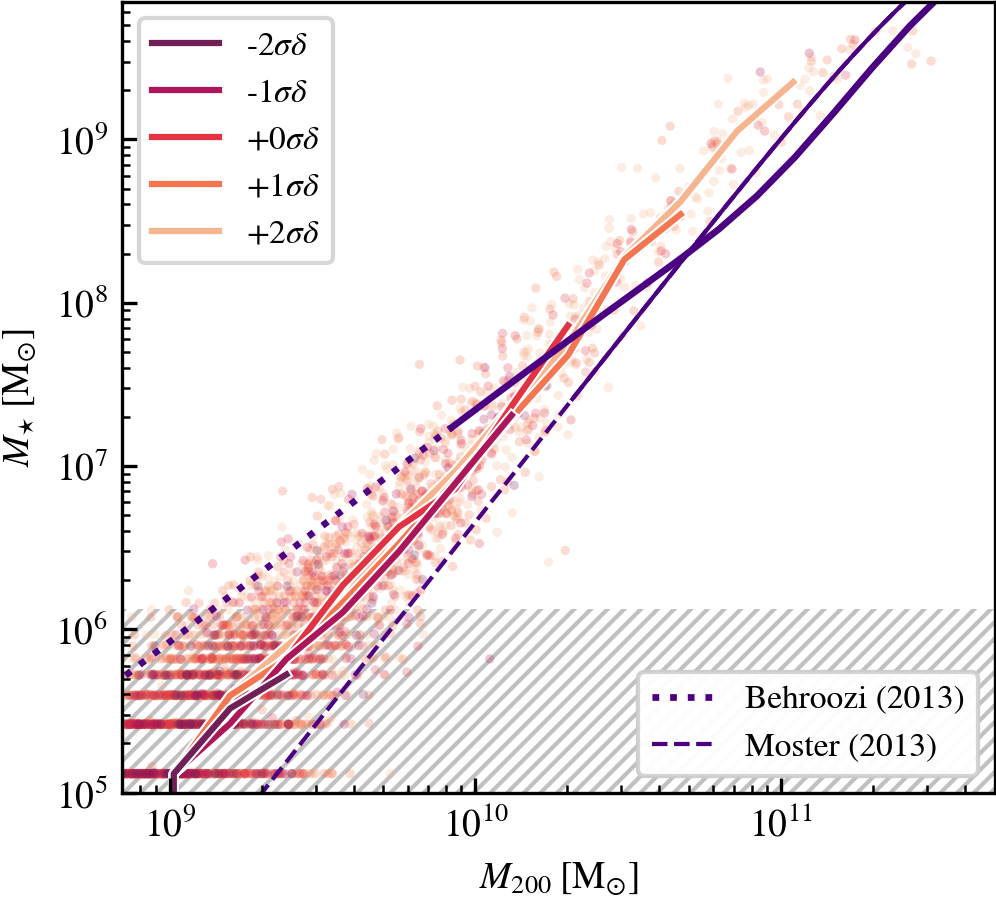}
    \caption{The present-day central stellar mass - halo mass scaling relation, at m5 resolution, of the aggregated central galaxy populations of the five regions comprising each of the five density bands. Coloured curves denote, for each density band, the median stellar mass at fixed halo mass, and include the contribution of haloes without a resolved galaxy. The SMHM relations of the semi-empirical models of \citet{Behroozi_2013} and \citet{Moster_2013} are drawn as dotted and dashed indigo curves, respectively, which transition to solid lines in the mass region to which they are fit.
    Colours, line styles and hatching follow the conventions introduced in Fig.~\ref{fig:GSMF}.}
    \label{fig:SHMR}
\end{figure}

Fig.~\ref{fig:SHMR} shows the present-day SMHM relation of central galaxies in the m5 simulations, with each coloured curve denoting the median central stellar mass at fixed halo mass for the aggregated galaxy population of the five regions comprising each density band, $\dfive$. The median is computed from all haloes, including those that do not host a galaxy (i.e. $M_\star = 0$), and is drawn for all halo bins sampled by at least 10 galaxies. Individual galaxies are also shown as filled symbols, colour coded according to the $\dfive$ of their parent region as per the median relation. The horizontal banding of individual galaxies in the low-mass regime is due to discrete numbers of stellar particles becoming visible on the logarithmic vertical scale.

Aside from the truncation of the curves due to the absence of massive haloes in lower density regions, the median SMHM relations of the regions comprising each of the $\dfive$ bands are remarkably similar, in both amplitude and slope. The scatter about the median stellar mass at fixed halo mass is also similar. Cosmological environment on the $r=5\,\cMpc$ scale defined by the $\dfive$ bands therefore clearly plays little role in shaping the halo-specific formation `efficiency' (if one defines this as the stellar mass of the central galaxy formed per unit halo mass) of isolated dwarf galaxies. The plot further underscores that the differing normalised stellar mass functions (shown in the bottom panel of Fig.~\ref{fig:GSMF}) are not a consequence of a systematic influence on this efficiency of the environment defined on this scale.

At first glance, our findings appear to contradict those of \citet{Zehavi2018}, who examined the influence of halo assembly bias on the galaxy population predicted by a semi-analytic model when applied to the Millennium Simulation \citep{Springel_MS_2005}, and those of \citet[][see also \citealt{Meshveliani_2024}]{Artale2018} who performed a similar study using the EAGLE Ref-L100N1504 simulation. In both cases, the authors report a significant correlation between the median stellar mass (at fixed halo mass) and the total mass density measured within a sphere of $r=5\,\hMpc$, for haloes of mass $\mhalo \gtrsim 10^{11} \Msun$. This halo mass range is effectively unsampled by our simulations (see Fig.~\ref{fig:HMF}) so our comparison does not probe the same fixed halo mass scale examined, nevertheless, we highlight two significant factors that likely contribute to the apparent difference in conclusions. 

Firstly, both studies were able to examine a significantly larger dynamic range in density (approximately $3\,{\rm dex}$ in the Millennium Simulation and $2\,{\rm dex}$ in EAGLE) than is the case here (approximately $0.8\,{\rm dex}$). The positive correlation they find between $M_\star(\mhalo)$ and the environmental density is particularly apparent as an elevated formation efficiency for galaxies located in high-density regions, which are absent from our sample by construction. Secondly, those studies measure the environmental density in a sphere centred on each galaxy, and thus perform a significantly different test to that presented in Fig.~\ref{fig:SHMR}. This is particularly true for low-density regions, since by definition their constituent galaxies tend to be located close to the boundary of the region, maximising the difference between the enclosed density of an arbitrarily-centred sphere, and one centred on a galaxy. 
More direct comparison to the findings of \citet{Zehavi2018} and \citet{Artale2018} can be achieved with halo specific properties, to which we turn to in Section\,\ref{sec:scatter_smhm}. 
Here we simply note that our simulations indicate an absence of systematic difference in the typical stellar mass of central dwarf galaxies at fixed halo mass in regions of differing enclosed density on a scale of $r=5\,\Mpc$. We cannot rule out an influence of environment, on larger or smaller scales, on dwarf galaxy evolution, though we note that \citet{Wu_2024} recently concluded from an analysis of the IllustrisTNG-300 simulation that the galaxy-halo connection is most significantly influenced by the environment measured on a scale of $\simeq 3\,\Mpc$. 

\subsection{Halo concentration as a function of halo mass}
\label{sec:results:concentration}

\begin{figure}
    \centering
    \includegraphics[width=\columnwidth]{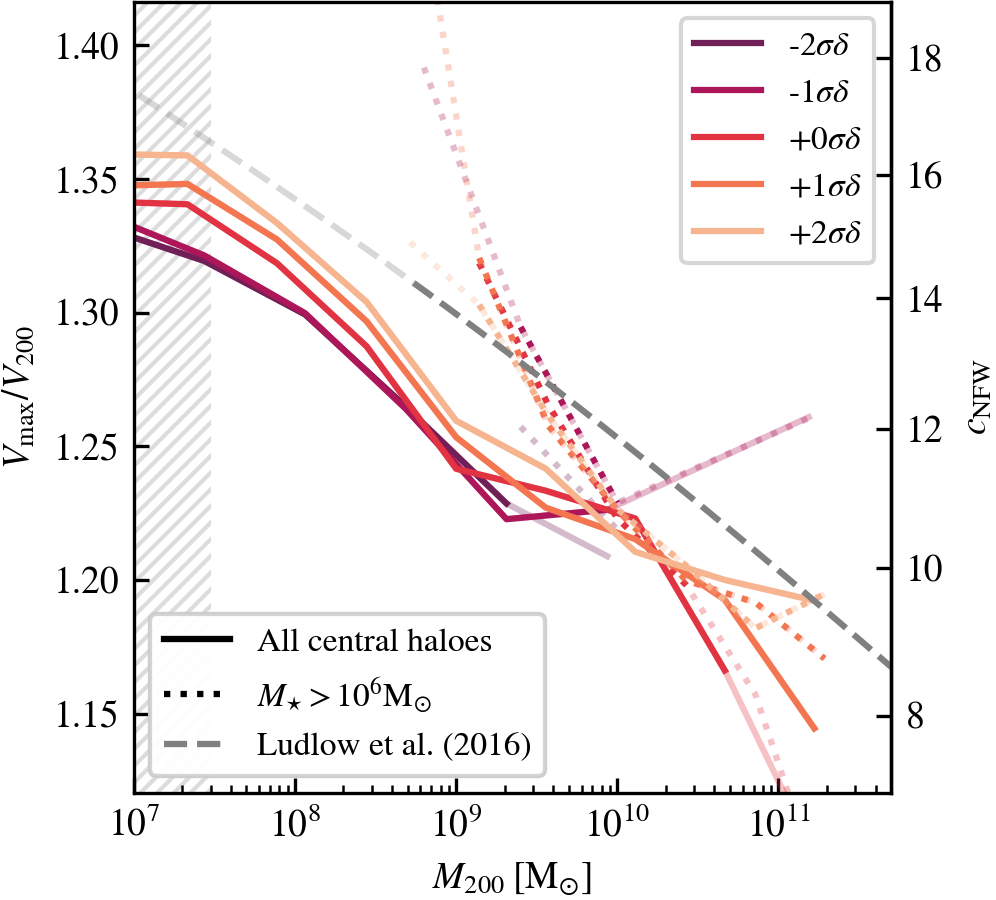}
    \caption{The present-day halo mass - concentration relation, at m5 resolution, of the aggregated field halo populations of the five regions comprising each of the five density bands. Solid coloured curves denote, for each density band, the median halo concentration at fixed halo mass, and include the contribution of haloes without a resolved galaxy. The dotted curves show the corresponding relation of only those haloes that host a galaxy of stellar mass $M_\star>10^6\,\Msun$. Halo concentrations are estimated from $\vmax/V_{\rm 200}$ ratios, as detailed in Sec. \ref{sec:concentration_calc}. The grey dashed line is the fit to the mass-concentration relation presented in \citet{Ludlow_2016} for systems above $\mhalo>5\times10^8\Msun$. Colours, line styles and hatching follow the conventions introduced in Fig.~\ref{fig:HMF}. There is a systematic shift in mass - concentration of halos between different overdensities, especially at the low mass end. Haloes hosting luminous galaxies are more concentrated than average.}
    \label{fig:m200_c}
\end{figure}

The intrinsic concentration of DM haloes is anti-correlated with their mass \citep[e.g.][]{Navarro_1996, Ludlow_2014, Correa_2015}. This `mass - concentration relation' arises due to the hierarchical nature of structure formation in CDM cosmogonies, with the characteristic density of a halo being proportional to the mean cosmic mass density at its initial collapse time. Low mass haloes therefore generally exhibit higher concentrations, reflecting their collapse at earlier cosmic epochs, making concentration a reasonable proxy for the formation time of a halo \citep{Wechsler_2002, Zhao_2009, Ludlow_2014, Correa_2015}. This implies that regions of differing enclosed density may exhibit halo mass-concentration relations that exhibit systematic offsets from the relation obtained from mean-density volumes.

Fig.~\ref{fig:m200_c} shows the present-day median halo concentration at fixed halo mass obtained from the m5 simulations. The relation is shown for the population of field haloes identified in the hydrodynamical simulations, but we remind the reader that (per Section \ref{sec:methods:galaxy_identification}) concentration is measured from each halo's matched counterpart in the corresponding collisionless simulation, to eliminate the influence of baryons on the inner structure of haloes. Each coloured curve denotes the mass-concentration relation of the galaxy population aggregated from the five simulations comprising each density band, $\dfive$, and is drawn with reduced opacity where the $\Delta\log_{10}M_{200}=0.2$ bin is sampled by fewer than 10 haloes. Dotted curves show the relation recovered from the subset of haloes that host a galaxy of stellar mass $M_{\star} > 10^6\,\rm M_{\odot}$, the threshold corresponding to $\simeq 10$ stellar particles, above which GSMFs of the simulations are reasonably converged (see Section~\ref{sec:validation:gsmf}).

Interestingly, we find that the subset of `luminous' low-mass haloes is strongly biased towards high concentrations, and hence these haloes exhibit a very steep mass-concentration relation. For our choice of stellar mass threshold used to define a `luminous' halo, the mass-concentration relations diverge below $\mhalo \simeq 3 \times 10^{9} \rm M_{\odot}$, which is the typical mass of haloes hosting galaxies of stellar mass $M_{\star} \simeq 10^6 \rm M_{\odot}$. Low-mass haloes that host a galaxy therefore have assembly histories that are strongly biased to early formation times. Despite there being a clear influence of environment on the mass-concentration relation of the overall halo population, no such effect is seen for the subset of haloes hosting a galaxy. This implies that halo occupancy is mostly influenced by its assembly history (for which we use concentration here as a proxy) rather than its large-scale environment. From a physical point of view, gas associated with low mass haloes is readily unbound by heating from the metagalactic UV/X-ray background during and subsequent to the epoch of HI reionisation. Those present-day low mass haloes whose main progenitor reached a greater mass prior to reionisation (i.e. those with an earlier formation time) can more readily form a significant mass of stars by the present day. A similar conclusion was reached by \citet{Fitts_2017} and \citet{Rey_2019} from analysis of the FIRE and EDGE simulations, respectively.

Our choice of the stellar mass threshold used to define what constitutes a `luminous' galaxy is clearly arbitrary, and follows from resolution considerations. For an alternative choice of threshold, the mass-concentration relations would diverge close to the halo mass that typically hosts galaxies of that stellar mass (see Fig.~\ref{fig:concentration_convergence}). Relatedly, it is clear that the steep negative gradient of the mass-concentration relation of luminous haloes can be attributed to the steep positive gradient of the low-mass regime of the SMHM relation (see Fig.~\ref{fig:SHMR_convergence}), as small changes in halo mass correspond to large changes in stellar mass. As is also clear from Fig.~\ref{fig:concentration_convergence}, the inferred mass-concentration relation of luminous haloes is insensitive to resolution, so long as one adopts a stellar mass threshold corresponding to at least 10 stellar particles.

\subsection{Halo luminous fraction as a function of mass}
\label{sec:results:lum_frac}

\begin{figure}
    \centering
    \includegraphics[width=\columnwidth]{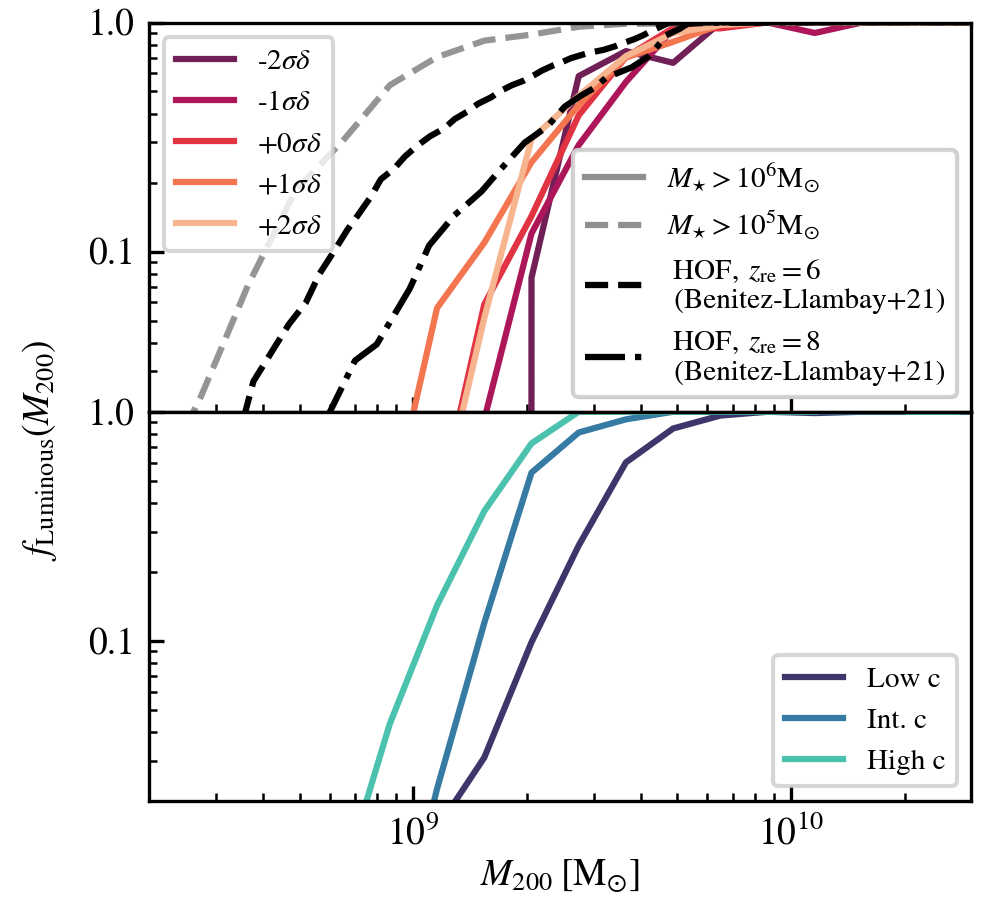}
    \caption{\textit{Upper panel:} The present-day luminous fraction as a function of mass, at m5 resolution, of the aggregated halo population of the five regions comprising the five density bands (coloured curves denote each density band). We deem haloes to be luminous if they host a galaxy with stellar mass $M_{\star}>10^{6}\,\Msun$. The luminous fraction is only weakly sensitive to large-scale environment. The grey dashed curve shows the luminous fraction of the aggregated halo population of all 25 simulations (i.e. not split by density band) recovered if one assumes haloes with galaxies $M_{\star}>10^{5}\,\Msun$ are luminous. The dashed and dot-dashed black curves show the halo occupation fractions of the semi-analytic model of \citet{BenitezLlambay2020}, assuming an HI reionisation redshift of $z=6$ and $z=8$.
    \textit{Lower panel:} The present-day luminous fraction of the total halo population of all 25 simulations at m5 resolution, with haloes binned into the low, intermediate, and high terciles of the intrinsic halo concentration distribution.}
    \label{fig:flum}
\end{figure}

An alternative way of viewing the divergence of the mass-concentration relations of dark and luminous haloes, is as a concentration dependence of the luminous fraction of haloes as a function of their mass. The upper panel of Fig.~\ref{fig:flum} shows the median luminous fraction (where luminous haloes are again defined as those hosting a galaxy of at least $M_\star = 10^6\,\Msun$) as a function of halo mass, for the aggregated galaxy populations corresponding to each density band, $\dfive$ at m5 resolution. The relations recovered from each density band are similar. All haloes with $\mhalo \gtrsim 10^{10}\,\Msun$ host a `luminous' galaxy, while there are no luminous galaxies in any halo of $\mhalo \lesssim 10^9\,\Msun$, with the luminous fraction showing a weak dependence on $\dfive$ close to $\mhalo \lesssim 10^9\,\Msun$. We attribute this difference to the systematic shift in the mass-concentration relation of all haloes at varying $\dfive$, apparent in Fig. \ref{fig:m200_c}. 

The luminous fraction smoothly transitions between $f_{\rm luminous} = 0$ and $1$, with approximately half of haloes with mass $\mhalo \simeq 3 \times 10^{9}\,\Msun$ being luminous. Unsurprisingly, this is the same mass below which the mass-concentration relation of luminous haloes begins to diverge from the relation of all haloes. As with the latter relation, the position of these curves shifts in response to an alternative choice of the stellar mass threshold used to define a luminous halo (see Fig. \ref{fig:flum_conv}). 

We caution that our definition of the luminous fraction is not equivalent to the absolute halo occupation fraction, and reiterate that our choice of stellar mass threshold ($M_{\star} = 10^6\,\Msun$) is guided by resolution considerations.
The luminous fraction based on this threshold should therefore be considered a lower limit. To highlight the sensitivity of the luminous fraction to a choice of a lower stellar mass fraction, we draw with a dashed grey curve the fraction of haloes, drawn from all 25 simulations at m5 resolution, with at least one stellar particle (broadly corresponding to a `galaxy' mass of $M_{\star} = 10^5\,\Msun$), whilst noting that this test is particularly sensitive to the stochasticity of the simulation's star formation implementation. We show this curve primarily for illustrative purposes, and to enable comparison with predictions for the absolute halo occupation fraction, which is an highly debated topic in the literature.

Two mass scales influencing the halo occupation fraction are often discussed. Photoheating during the epoch of reionisation, and the subsequent presence of a metagalactic UV/X-ray background, unbinds gas from haloes of present-day mass $\mhalo \lesssim 10^{9.5}\,\Msun$, suppressing the occupation fraction of less massive haloes below unity. The other mass scale is that corresponding to a halo virial temperature comparable to the hydrogen atomic cooling limit ($T \simeq 7000\,{\rm K}$), which is relevant before the epoch of reionisation. This scale is redshift dependent, $M \propto (1+z)^{-3/2}$, and at the present day is $\mhalo \sim 10^9\,\Msun$. Gas associated with haloes of mass below this limit cannot efficiently cool and form stars. A large fraction of haloes with present-day mass $\mhalo < 10^{8.5}\,\Msun$ never crossed the atomic hydrogen cooling limit during their formation history and are often predicted to be devoid of stars at the present day. \citet[][hereafter BLF20]{BenitezLlambay2020} construct a semi-analytical model based on these mass scales and realistic halo growth histories to estimate the absolute halo occupation fraction, with cooling, the UV background, and reionisation assumptions that differ from those in our simulations. The black dashed and dashed-dotted curves in the top panel of Fig.~\ref{fig:flum} show the halo occupation fraction of this model when the epoch of reionisation is assumed to be $z_{\rm reion}=6$ and 8, respectively, encompassing what is assumed in our simulations ($z_{\rm reion}=7.5$). Despite our simulations' necessary limitation of a minimum mass threshold of $M_{\star} > 10^5\,\Msun$ (which renders this prediction somewhat of a lower limit for the true occupation fraction), the model of BLF20 predicts noticeably lower occupation fractions than our simulations. This mismatch can be attributed to a number of differences between the two models. The model of BLF20 assumes gas is in hydrostatic equilibrium within spherical dark matter haloes of mean concentration, and that the gas contributes negligibly to the gravitational potential at all radii, i.e. gas does not become self-gravitating. Moreover, the gas in the centre of haloes in their model is assumed to be isothermal at $T\sim10^4$K (no cold phase). Other potentially relevant differences are the star formation criteria, assumed cooling rates, and the amplitude of the background UV ionising radiation. Identifying the source of the difference between our prediction and that of BLF20 in more detail is beyond the scope of this paper and an interesting topic for future work.

The weak dependence of the mass-concentration relation on $\dfive$ for luminous haloes seen in Fig.~\ref{fig:m200_c} indicates that internal halo structure (i.e. assembly-bias related effects) is a more important factor than large-scale environment in governing the stellar mass of haloes. To test this explicitly, in the lower panel of Fig.~\ref{fig:flum} we revert to our standard threshold of $M_{\star} = 10^6\,\Msun$, and show the median luminous fraction of haloes drawn from all 25 simulations, binned into the lower, central and upper terciles of the intrinsic halo concentration distribution: high concentration haloes have $ \cnfw > 18.1$, intermediate concentration haloes have $12.7< \cnfw <18.1$, and low concentration haloes have $\cnfw<12.7$. The luminous fraction changes significantly and systematically with halo concentration. For example, at $\mhalo = 2 \times 10^9 \Msun$, the occupation fraction changes from $\simeq 10$ percent for low concentration haloes to $\simeq 80$ percent for high concentration haloes. The halo masses corresponding to 50 percent occupancy for low, intermediate and high concentration bins are $M_{\rm 50}=3.2\times10^9 \Msun$,  $2.1\times10^9 \Msun$, and $1.7\times10^9 \Msun$, respectively. 
These results further underscore that the formation and evolution of low-mass dwarf galaxies are closely tied to their halo assembly history, rather than the large scale environment.

\section{Scatter in the stellar mass - halo mass relation}
\label{sec:scatter_smhm}

\begin{figure}
    \centering
    \includegraphics[width=\columnwidth]{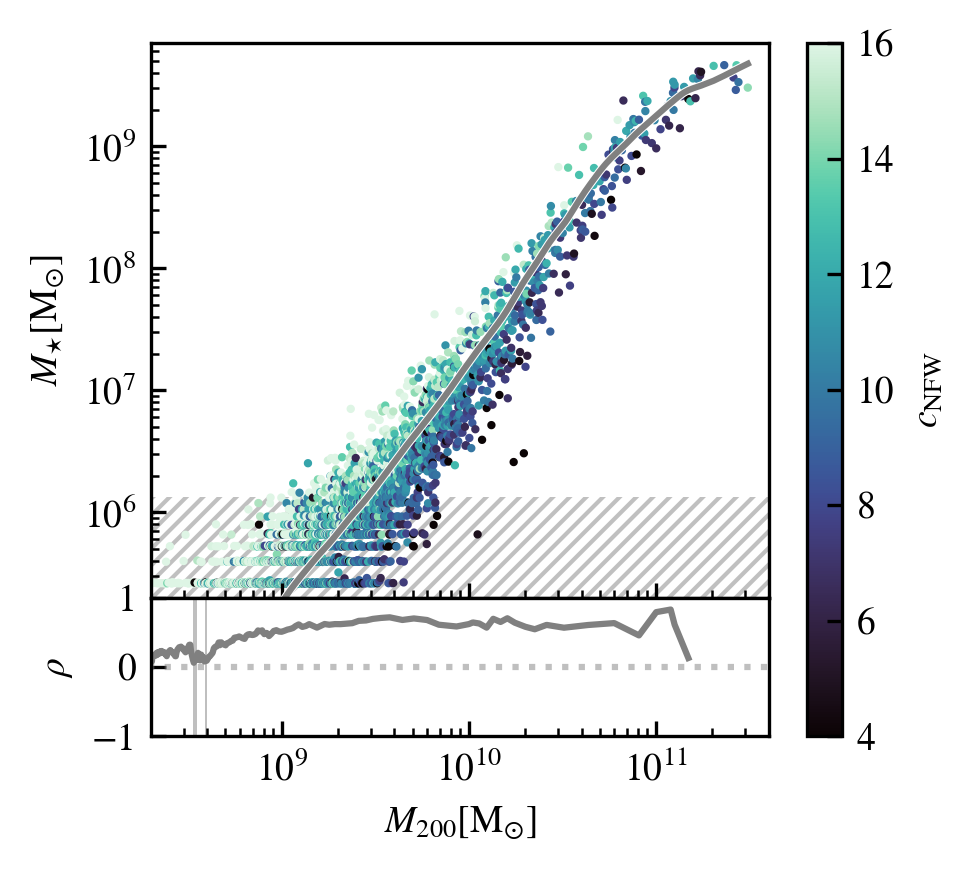}
    \caption{\textit{Top}: The stellar mass versus halo mass ($\mhalo)$ for field galaxies in our 25 simulations at m5 resolution. Each symbol represents an individual galaxy and is coloured by the halo concentration, $\cnfw$, of their dark matter hosts, measured from the $V_{\rm max}/V_{200}$ ratio in the counterpart collisionless simulations. The grey curve shows the median relation computed using the LOWESS method. The scatter at a given halo mass shows a clear trend with concentration, with higher stellar masses in more concentrated haloes. The grey hatched region indicates where galaxies have fewer than 10 star particles. 
    \textit{Bottom}: The running Spearman rank correlation coefficient, $\rho$, between the residuals $\Delta c_{\rm NFW}$ and $\Delta M_{\star}$. Positive values close to 1 shows a strong positive correlation for all well resolved objects. The shaded regions indicate when the correlation has low significance (p>0.01).
    }
    \label{fig:SMHM_c_lowess_combined}
\end{figure}

The findings presented in the previous section highlight that the influence of the large-scale ($\simeq 5\,\Mpc$) environment on the properties of low-mass central galaxies is driven primarily from differences in the overall halo population (such as the distributions of halo concentration or halo formation time) that emerge in differing environments. In this section, we therefore examine how these key halo properties that mediate the environmental differences correlate with the scatter of the SMHM relation. Section \ref{sec:scatter_smhm:conc} focuses on halo concentration and Section \ref{sec:scatter_smhm:tform} on galaxy formation times.

\subsection{Correlation with halo concentration}
\label{sec:scatter_smhm:conc}

Fig.~\ref{fig:SMHM_c_lowess_combined} shows the SMHM relation of the composite galaxy population drawn from all 25 regions at m5 resolution. Symbol colour denotes the intrinsic concentration of the halo, $\cnfw$, and the solid grey curve shows the running median of $M_{\star}$ as a function of $\mhalo$, computed using the locally-weighted scatter plot smoothing method \citep[LOWESS,][]{Cleveland_79}. We use this method rather than a binned median as it enables the calculation of a median value corresponding to the precise mass of each halo. In the range $M_{\star}= [10^6-10^9] \Msun$, this curve is well fit by a power-law of form $M_{\star} /\Msun = \beta (M_{\rm 200} /10^{10}\Msun)^{\alpha}$ where $(\alpha,\beta)=(2.17,1.42\times 10^7)$, as shown in Appendix~\ref{appendix:fit}.

Scatter about the SMHM relation at fixed halo mass is strongly and positively correlated with halo concentration such that more concentrated haloes form more massive galaxies. This relation has previously been reported elsewhere for large samples of more massive ($\mhalo \gtrsim 10^{11}\,\Msun$) simulated galaxies \citep[e.g.][see also \citealt{Montero-Dorta_2020, Wang_2025}]{Matthee_2016}. To quantify the strength of the correlation as a function of halo mass, we compute for each galaxy, $i$, the offset between the value of $M_{\star,i}$ and the LOWESS median of this quantity at $M_{200,i}$, which we define as $\Delta M_{\star, i}$. We then compute the equivalent metric for halo concentration, $\Delta c_{{\rm NFW},i}$, and compute the Spearman rank correlation coefficient, $\rho$, of the two quantities in samples defined by a `rolling' window in halo mass. This quantity is plotted in the subpanel of Fig.~\ref{fig:SMHM_c_lowess_combined}. Solid shading in the panel corresponds to halo mass windows in which the probability of a chance correlation is high, denoted by Spearman $p$-value greater than 0.01. A significant positive correlation between stellar mass and halo concentration at fixed halo mass is present for $M_{200} \gtrsim 5\times10^8 \Msun$, though we note that for $M_{\rm 200} \lesssim 5\times10^9 \Msun$, the finite resolution of the simulation hinders the sampling of the stellar mass, as is clear from the diagonal hatching in the main panel of the plot.

\citet{Matthee_2016} attributed the correlation between stellar mass and halo concentration at fixed halo mass to a combination of earlier halo assembly and the increased gravitational binding energy of more concentrated haloes, which reduces the (macroscopic) efficiency of feedback. More recently, \citet{Wang_2025} argue that, for haloes with $\mhalo\approx10^{11}-10^{12}\Msun$, variations in binding energy are the dominant driver of this correlation.
As also noted by \citet{Fitts_2017}, at the low halo masses explored here ($M_{200} \lesssim 10^{10}\,\Msun$) photoheating during and after reionisation also has a significant influence on the halo's star formation history. The earlier assembly of more concentrated haloes affords them an elevated probability of forming stars prior to the epoch of reionisation, and owing to their greater binding energy they lose a smaller fraction of their gas to photoheating. The scatter about the SMHM relation in the simulations increases towards lower halo masses, becoming particularly large below $M_{200} \lesssim 10^{10}\,\Msun$. Quantified in $0.2 \,{\rm dex}$ bins, the 10th-90th percentile scatter in $M_{\star}$ at fixed $M_{200}$ increases from $0.71 \,{\rm dex}$ at $M_{200}=1.7\times10^{10}\,\Msun$ to $0.85 \,{\rm dex}$ at $M_{200}=4.5\times10^9\,\Msun$. We speculate that this may be a signature of the role played by reionisation in exacerbating the dependence of $M_{\star}/M_{200}$ on halo concentration and formation time.

\begin{figure}
    \centering
    \includegraphics[width=\columnwidth]{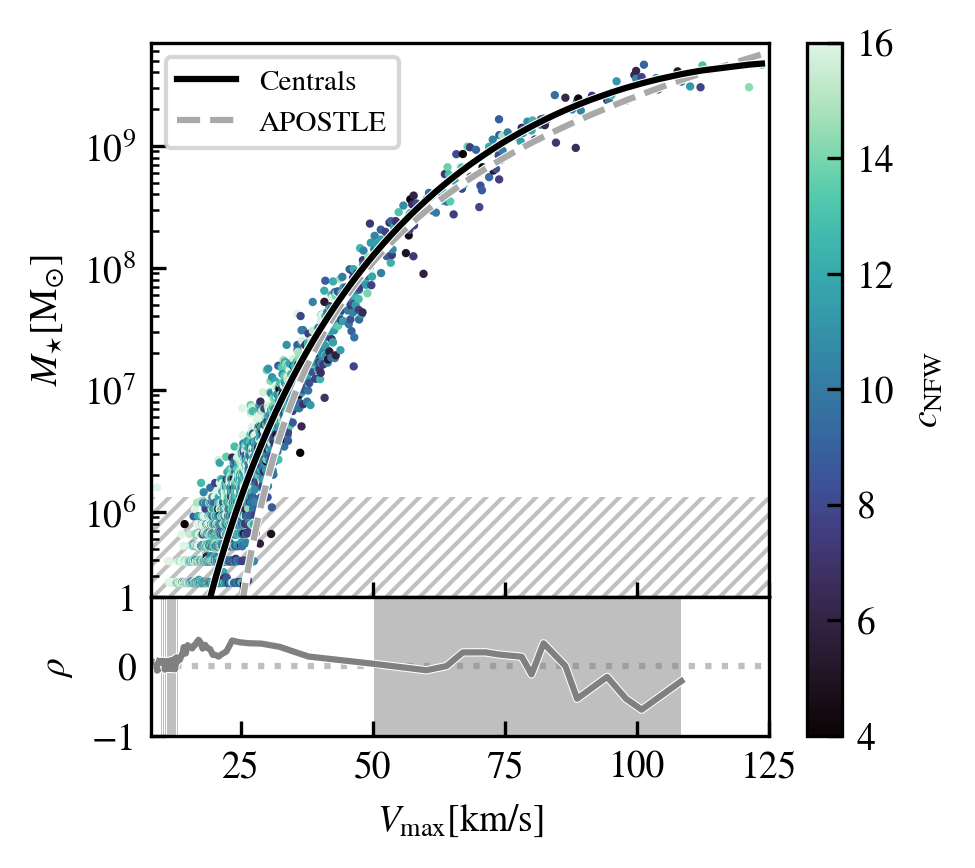}
    \caption{The same as Fig.~\ref{fig:SMHM_c_lowess_combined} but showing $V_{\rm max}$ instead of halo mass. The scatter in stellar mass is smaller, compared to the SMHM relation of Fig.~\ref{fig:SMHM_c_lowess_combined}. Moreover, unlike the $M_{\star}-M_{\rm 200}$ relation, the scatter in the $M_{\star}-V_{\rm max}$ relation shows no clear correlation with concentration. The running Spearman rank correlation coefficient is shown in the bottom panel. A fit to the median relation is shown in black, following the form $M_{\star}/\Msun = m_0 \nu^{\alpha} exp (-\nu^{\gamma})$ where $\nu=\vmax/50$km/s and ($m_0,\alpha,\gamma$)=($3.39\times10^8$, 7.56, 1.59). A fit to the APOSTLE simulation \citep{Fattahi_2018} data is shown by the dashed grey curve.}
    \label{fig:SMVR_c_lowess}
\end{figure}

It has long been argued that stellar mass and luminosity should correlate more strongly with the maximum value of a halo's circular velocity profile than with its halo mass \citep[e.g.][]{Nagai_2005,Conroy_2006}, as the maximum circular velocity more directly traces the depth of the central potential well, and hence the ability to retain gas against feedback-driven outflows, independently of the halo's outer mass distribution. \citet{Matthee_2016}, using the EAGLE simulations, and \citet{Fitts_2017}, using 15 zoom-in simulations of dwarf galaxies from the FIRE simulations, found this to indeed be the case. We therefore show in Fig.~\ref{fig:SMVR_c_lowess} the stellar mass, $M_{\star}$, of field galaxies drawn from the composite galaxy population as a function of the maximum value of their total circular velocity profile, $\vmax$. Symbol colour denotes the halo concentration, $\cnfw$ and, as in Fig. \ref{fig:SMHM_c_lowess_combined}, the solid black curve shows the median $M_{\star}-\vmax$ relation computed using the LOWESS method, which can be fit using the function $M_{\star} [\Msun] = m_0\,\nu^{\alpha}\,exp{(-\nu^{\gamma})}$ where $\nu=\vmax/50\,{\rm km\,s}^{-1}$ and ($m_0,\alpha,\gamma$)=($3.39\times10^8\,\Msun$, 7.56, 1.59). The dashed grey curve shows the corresponding relation inferred from the APOSTLE simulations \citep{Fattahi_2018}. The two suites yield a similar outcome: over the range $30 < \vmax < 100\, \kms$, the largest difference between the median values of $\mstar$ at fixed $\vmax$ is $0.41 \,{\rm dex}$, which is found at the low mass end where the relation is steep. 

As expected, the scatter in $\mstar$ at fixed $V_{\rm max}$ is far smaller than at fixed $\mhalo$, with the difference most-readily quantified via the scatter in the $x$-axis quantity at fixed $\mstar$: in the interval $10^{7}<\mstar/\Msun<10^{8}$, the 10$^{\rm th}$-90$^{\rm th}$ percentile scatter at fixed $V_{\rm max}$ is $0.14\,{\rm dex}$, whilst the corresponding value at fixed $\mhalo$ is $0.47\,{\rm dex}$. The sub-panel of Fig.~\ref{fig:SMVR_c_lowess} shows that only a small residual correlation between $\mstar$ and $\cnfw$ at fixed $\vmax$ remains, and then only for resolved haloes with $\vmax \lesssim 50\,{\rm km\,s}^{-1}$. As previously demonstrated by \citet{Matthee_2016} at larger halo mass scales, halo concentration (or, equivalently, halo binding energy) accounts for the bulk, but not all, of the scatter about the SMHM relation.

\subsection{Correlation with galaxy formation time}
\label{sec:scatter_smhm:tform}

\begin{figure*}
    \includegraphics[width=0.75\textwidth]{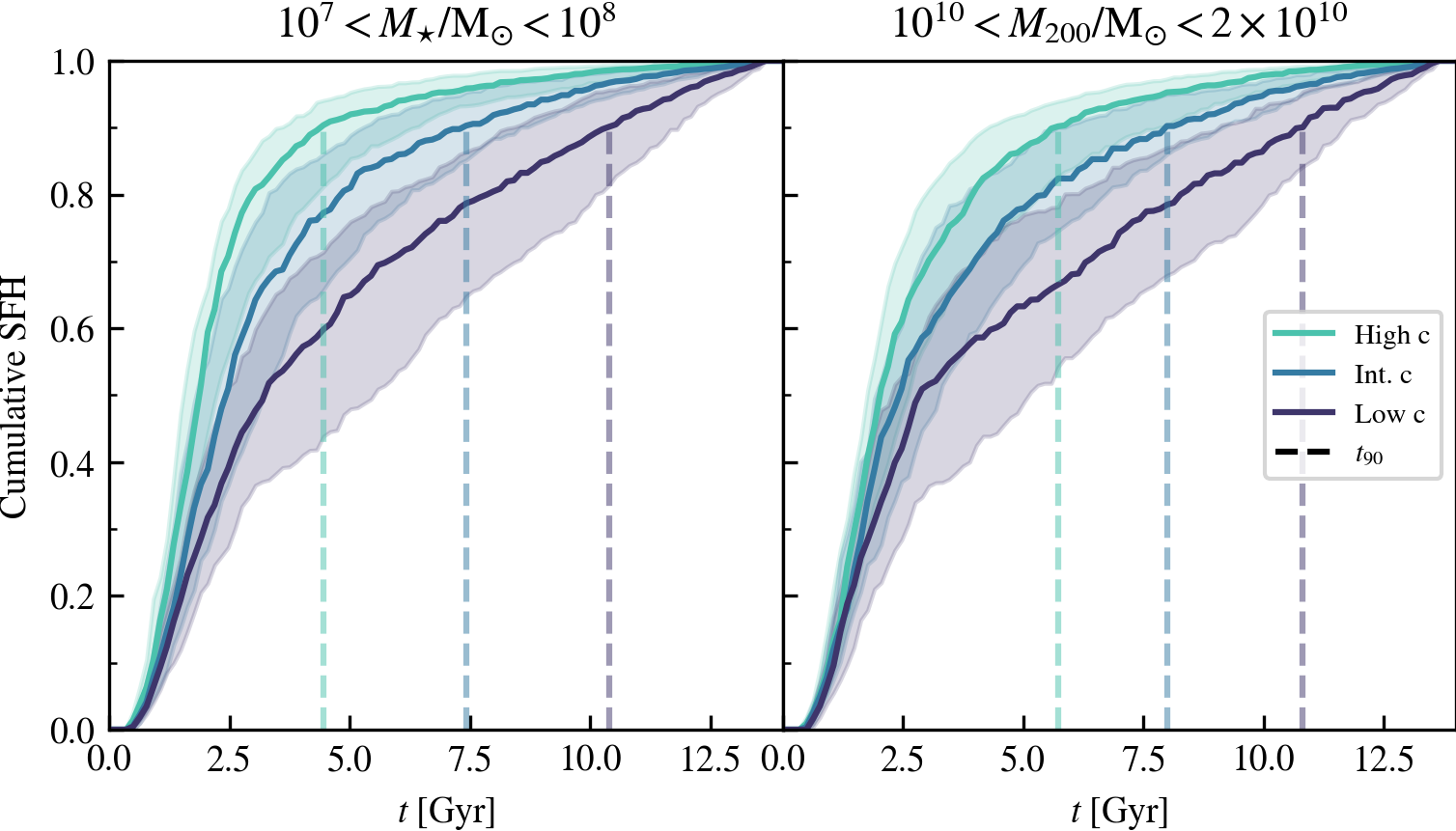}
    \caption{The cumulative star formation histories (SFH) of central galaxies in the simulation suite, divided by their stellar mass at $z=0$ and their dark matter halo concentration. The left panel corresponds to a present-day stellar mass bin ranging between $10^{7}<M_{\star}/ \Msun<10^{8}$ and the right panel corresponds to a present-day halo mass bin of $10^{10}<M_{200}/ \Msun <2\times10^{10}$. In both panels, the sample is split into three based on the terciles of the halo concentration distribution. The cumulative SFHs are normalised to the $z=0$ stellar mass first, and then median and scatter (25th-75th percentiles) are computed and shown by the solid lines and shaded regions, respectively.
    The median times at which 90\% of the final stellar mass has formed, $t_{90}$, are shown by the vertical dashed lines. Consistent over the full mass range, higher concentration galaxies have earlier SF, reflected by earlier $t_{90}$. Time and $t_{90}$ correspond to cosmic time.}
    \label{fig:SFH_c}
\end{figure*}

Having established that the scatter about the SMHM relation in the low-mass regime is strongly and positively correlated with halo concentration, a corollary is that the scatter might also correlate with the galaxy's star formation history (SFH), owing to the correlation between halo concentration and halo assembly history. We therefore explore the influence of halo concentration on the star formation histories of our simulated galaxies by showing the cumulative, normalised SFH of galaxies with present-day stellar mass of $10^7< \mstar/\Msun < 10^8$ in the left panel of Fig.~\ref{fig:SFH_c}, binned into terciles of present-day intrinsic halo concentration. The solid curves and shaded regions denote the median and interquartile range of the cumulative SFHs. There is a significant systematic trend such that galaxies hosted by more concentrated haloes typically form their stars earlier. The difference in SFHs is not only present at early times, but persists throughout the assembly of the galaxies; e.g. the cosmic time by which 90 percent of present-day stellar mass has formed, $t_{90}$, is $4.4\,{\rm Gyr}$, $7.4\,{\rm Gyr}$ and $10.4\,{\rm Gyr}$ for the median SFH of galaxies in the high, intermediate and low concentration samples, respectively. These times are denoted by vertical dashed lines on the figure.

Since halo mass and concentration correlate, we isolate the influence of concentration alone on the previous trend by showing in the right panel of Fig.~\ref{fig:SFH_c} the SFHs of galaxies hosted by haloes within a narrow range of halo mass, $10^{10} < M_{200}/\Msun < 2\times10^{10}$. The trend between SFH and halo concentration is still evident: high concentration haloes, i.e. earlier-forming haloes host earlier-forming galaxies. The average $t_{90}$, indicated by the vertical dashed lines, shows a systematic shift to earlier cosmic times as the halo concentrations increase, corresponding to $5.7\,{\rm Gyr}$, $8.0\,{\rm Gyr}$ and $10.8\,{\rm Gyr}$ for the high, intermediate, and low concentration bins respectively. These results emphasize the link between halo assembly and stellar assembly times.

\begin{figure}
    \centering
    \includegraphics[width=\columnwidth]{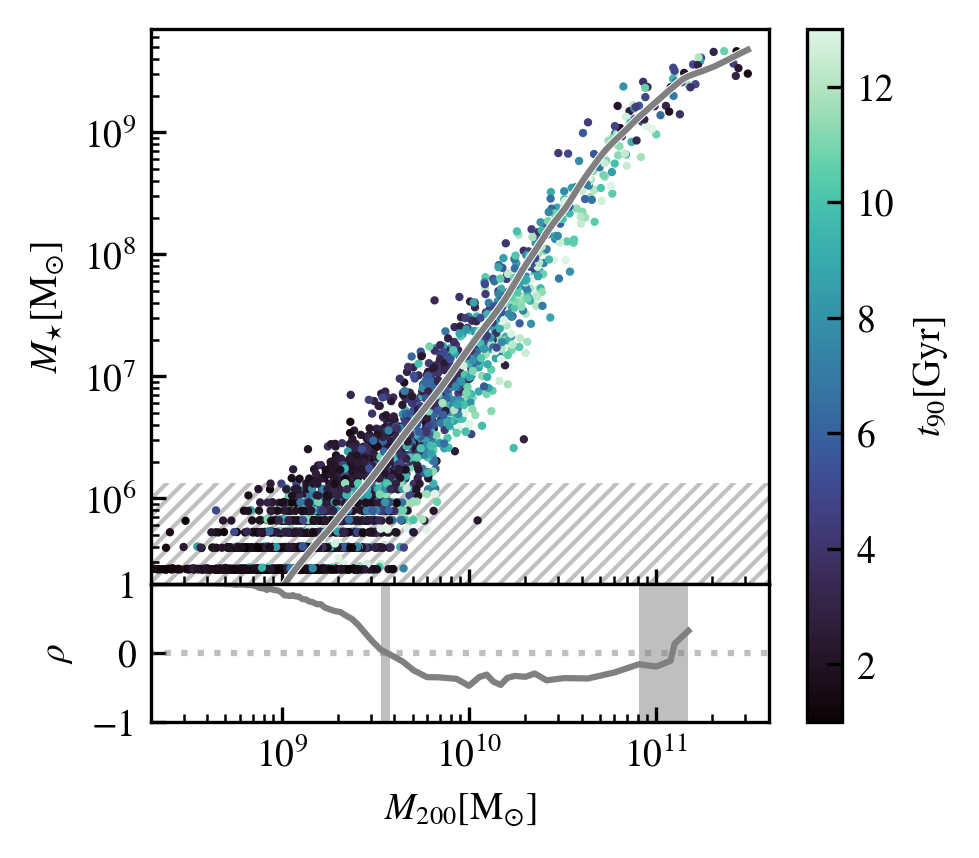}
    \caption{Similar to Fig. \ref{fig:SMHM_c_lowess_combined} but points are colour-coded according to $t_{90}$, the cosmic time at which 90\% of the total stellar mass of a galaxy has formed.} 
    \label{fig:SHMR_tau90_lowess_combined}
\end{figure}

The previous finding indicates a likely correlation between SFH and the scatter about the SMHM relation since, as shown in Sec. \ref{sec:scatter_smhm}, the latter correlates with halo concentration. We explore this in Fig.~\ref{fig:SHMR_tau90_lowess_combined} where we repeat Fig. \ref{fig:SMHM_c_lowess_combined}, but here colour code the symbols by $t_{90}$.
In the regime where galaxy masses, and hence SFHs, are well sampled, there is an evident anti-correlation between the scatter about the SMHM relation and $t_{90}$\footnote{Our results are qualitatively similar if using $t_{10}$ or $t_{50}$.}. At fixed halo mass, galaxies with greater stellar mass have lower $t_{90}$ and form their stars earlier. The running Spearman rank correlation coefficient between the residuals of stellar mass, $\Delta M_{\star,i}$ and $t_{90}$, $\Delta t_{90,i}$, is shown as a function of halo mass in the bottom panel of Fig. \ref{fig:SHMR_tau90_lowess_combined}, which highlights the anti-correlation ($\rho<0$) for the halo mass interval $ 4\times10^9 \lesssim M_{200}/\Msun \lesssim 10^{11}$. At greater masses, SFHs are influenced by AGN feedback and the role of concentration is more complicated in such galaxies because it can accelerate the growth of the central BH \citep[e.g.][]{Davies2019}.

The aforementioned anti-correlation between $M_{\star}$ and $t_{90}$ may appear to contradict the expectation that higher mass dwarf galaxies have more prolonged star formation and hence larger $t_{90}$ \citep[see e.g.,][]{Simpson_2018,Digby_2019}. However, we emphasize that the anti-correlation applies when considering galaxies at fixed halo mass, and does not apply to the average $M_{\star}-t_{90}$ relation where there is a clear overall trend such that more massive galaxies have a later formation time, except for those galaxies hosted by haloes of mass $\mhalo \gtrsim 10^{11}\,\Msun$ whose properties are not accurately reproduced by these simulations (see Section \ref{sec:validation}). The positive correlation between $M_\star$ and $t_{90}$ is evident in Fig. \ref{fig:mstar_t90}.

The star formation history of dwarf galaxies, especially at late cosmic times, is more readily inferred from observations than their halo concentration. Thus, the correlation of the scatter about the SMHM relation and the $t_{90}-M_\star$ relation provides a test of our galaxy formation model, one that can, in principle, be corroborated or falsified by observations. We note that the predicted shape of, and scatter about, the SMHM relation are sensitive to the details of the adopted galaxy formation model, but we anticipate that the correlations are reasonably robust to such details because they link properties of DM haloes (halo assembly and concentration) to the coarse-grained stellar assembly history. Even though we do not provide detailed comparison of SFHs or colour of our simulated dwarfs with observations, as it goes beyond the scope of the paper, we emphasize that the trends are expected to be robust.
Indeed, \citet{Kim2024} report that in the EDGE simulations the halo mass is more strongly correlated with the average star formation rates of dwarfs than with their absolute stellar mass. This is also motivated by observational constraints, as argued by \citet{Read_Erkal_2019} for dwarf galaxies in the Local Group.

%% file: Summary/summary.tex
\section{Summary and discussion}
\label{sec:summary}

We have introduced Columba, a suite of cosmological hydrodynamical simulations of the $\Lambda$CDM cosmogony that follow the formation and evolution of isolated dwarf galaxies in diverse low-density environments. The suite is comprised of 25 zoom-in simulations that follow the evolution of spherical regions with radius $r=5\,\Mpc$ at $z=0$, drawn from a collisionless simulation of a periodic volume with side length $L=400\,\cMpc$. The regions, which span diverse cosmic environments from voids to filament-like structures, are devoid of massive haloes with present-day mass comparable to or larger than that of the Milky Way. The average density of the target regions at $z=0$ corresponds to the (-2, -1, 0, +1, +2)$\sigma$ percentiles of the distribution of the mass density of present-day  $r=5\,\Mpc$ spheres that are devoid of massive halos. The corresponding overdensities are $\log_{10}\dfive = (-1.064, -0.859, -0.645, -0.449, -0.282)$, where $\delta_5 = \rho_5 / \bar{\rho}$,\, $\rho_5$ is the mean density within a sphere of radius $r=5\,\Mpc$ and $\bar{\rho}$ is the mean cosmic mass density. The five regions selected at each density, $\delta_5$, were chosen to further sample the distribution of the number of haloes with mass broadly corresponding to that of the Large Magellanic Cloud, $N_{\rm LMC}$, at the corresponding density.

The simulations were evolved to the present day with the cosmology, gravity, hydrodynamics and galaxy formation code \textsc{Swift} \citep{Schaller2024swift}, with approximately equal mass DM and gas particles, at two levels of resolutions. The first, `m6', corresponds to a gas particle mass of $m_{\rm gas} = 1.8 \times 10^6\,\Msun$, while the flagship `m5' resolution corresponds to $m_{\rm gas} = 2.3 \times 10^5\,\Msun$. The simulations adopt a galaxy formation model that incorporates substantial improvements over EAGLE and shares several key components with the COLIBRE model \citep{Schaye_COLIBRE, Chaikin_COLIBRE}. These include an explicit model of cold, dense neutral gas, a star formation model that accounts for gas turbulence, a treatment of pre-supernova feedback from massive stars, more sophisticated methods of distributing the energy liberated by SNe and active galactic nuclei (AGN), and updated treatments of BH growth and dynamics. While Columba inherits several modules from COLIBRE, it does not include a live dust model coupled to radiative cooling, a non-equilibrium H/He network, turbulent diffusion, updated chemical enrichment and SNIa schemes, or the jet-mode AGN feedback channel. The underlying subgrid prescriptions inherit parameter choices that result from testing performed for the EAGLE \citep{Crain_2015} and COLIBRE \citep{Chaikin_COLIBRE} simulation campaigns. As in all cosmological simulations, additional implementation choices such as the feedback heating temperature, and choice of particles selected to be directly influenced by feedback processes also indirectly influence galaxy properties. Unlike COLIBRE, we also do not vary the subgrid efficiency of feedback from SNe with resolution. Nevertheless, when rescaled to account for their preferential sampling of underdense regions, the simulations reproduce the present-day galaxy stellar mass function (GSMF) in the stellar mass regime they resolve and sample very well, $\mstar=10^6-10^9 \Msun$. The galaxies are slightly larger than their similar-mass counterparts observed in the Local Group, by approximately 60 per cent for the m5 resolution simulations. 

We investigate the influence of environment, characterized by $\dfive$, on the population of simulated dwarf galaxies. Our main findings are as follows:

\begin{itemize}
    \item Once the difference in the mean densities of the regions is accounted for, the low-mass power-law regime of the halo mass functions (HMFs) of the simulated regions are very similar, as shown in the bottom panel of Fig.~\ref{fig:HMF}. There is a mild second-order effect, such that the density-normalised space-density of haloes at fixed mass  ($\phi(M)/\rho_5$) is greater in \textit{lower} density regions. The characteristic mass scale at which the HMFs transition from a power law to an exponential decline changes systematically with $\dfive$, such that lower density regions host fewer high-mass haloes. This characteristic mass scale is $M_{200} \sim 10^9 \Msun$ in the lowest density regions, affecting the space density of haloes that host dwarf galaxies. 
    
    \item The HMFs are mildly suppressed compared to their collisionless (``DM-only'') counterparts in the low-mass regime as illustrated in the middle panel of Fig.~\ref{fig:HMF}. The suppression of the HMF is insensitive to the overdensity of the region; i.e.  $\phi_{\rm hydro}/\phi_{\rm DMO} \simeq 0.8$, for a halo mass of $\mhalo<10^9 \Msun$, for all environments.   

    \item Similar to the HMFs, Fig.~\ref{fig:GSMF} shows that the galaxy stellar mass function (GSMF) of regions with different mean density differ in the low-mass regime predominantly in their normalisation. The lowest density regions have systematically lower density-normalised GSMFs ($\phi(M_{\star})/\rho_5$) at all resolved stellar masses. We attribute this difference primarily to the dependence of the HMF break scale on $\dfive$.

    \item The median stellar mass-halo mass (SMHM) relation over the mass scales examined here can be represented by a power-law of form $M_{\star} /\Msun = \beta (M_{\rm 200} /10^{10}\Msun)^{\alpha}$ where $(\alpha,\beta)=(2.17,1.42\times 10^7)$, fitted to the range $M_{\star}= [10^6-10^9] \Msun$. Fig.~\ref{fig:SHMR} shows that we find no significant correlation of the median SMHM relation and $\dfive$.

    \item Fig.~\ref{fig:m200_c} shows that the median halo concentration at fixed halo mass is mildly correlated with $\dfive$, consistent with the trend that haloes in denser regions tend to have higher concentrations \citep[see also][]{Hellwing_2021}. However, the median halo concentration of `luminous' haloes (those hosting a galaxy of $\mstar > 10^6\Msun$ at m5 resolution) is insensitive to $\dfive$. The mass - concentration relation of this population diverges steeply from that of all haloes for $\mhalo \lesssim 3\times10^9 \Msun$ towards higher concentrations.  

    \item The fraction of luminous haloes, displayed in Fig.~\ref{fig:flum}, shows a small ($<30$ per cent between the regions of highest and lowest $\dfive$) but systematic correlation with $\dfive$, such that the luminous fraction at fixed mass is greater in higher-density regions. We attribute this trend to a strong dependence of the luminous fraction on the intrinsic DM halo concentration, $c_{\rm NFW}$. For example, for haloes of $M_{200}\simeq 2\times10^9\Msun$, the luminous fraction differs by $63$ percent between the upper and lower concentration terciles.
    
\end{itemize} 

Our study highlights the important and closely linked roles of halo assembly and halo concentration on the formation and evolution of dwarf galaxies. We have shown in Fig.~\ref{fig:SMHM_c_lowess_combined} that in the low-mass regime ($\mhalo=10^9-10^{11} \Msun$) the scatter in the $\mstar-\mhalo$ relation correlates strongly and positively with the concentration of a galaxy's host DM halo, when measured using the $V_{\rm max}/V_{200}$ ratio from the matched DM-only counterparts. A corollary of this correlation is that there is less scatter about the $\mstar-\vmax$ relation (Fig.~\ref{fig:SMVR_c_lowess}) than about the $\mstar-\mhalo$ relation. Scatter about the former relation does not correlate significantly with DM halo concentration. These findings are consistent with those of previous studies examining higher resolution simulations of small samples of dwarf galaxies \citep[e.g.][]{Fitts_2017, Rey_2019}. 

Studies based on hydrodynamical and/or semi-analytic simulations of large periodic volumes have previously revealed the strong correlation of scatter about the SMHM relation with the formation time or concentration of galaxy host DM haloes on relatively high stellar mass scales \citep[$M_{\rm halo} \gtrsim 10^{11} \Msun$, e.g.][]{Matthee_2016, Zehavi2018, Artale2018, Wang_2025}. The large samples yielded by our simulations demonstrate that these trends extend into the dwarf galaxy regime as shown in Fig.~\ref{fig:SMHM_c_lowess_combined}. 

In Fig.~\ref{fig:SFH_c} we show that the star formation histories of our simulated dwarf galaxies, either at fixed halo mass or fixed stellar mass, correlate strongly with halo concentrations. Lower concentration haloes, which typically form later, have star formation histories delayed by $\approx 5\,$Gyr in comparison to their high concentration counterparts. This link between star formation history and halo concentration (halo assembly) leaves its imprint in the scatter of the SMHM relation. The scatter of the SMHM relation at fixed halo mass is negatively correlated with $t_{90}$; i.e. higher stellar mass galaxies form the majority of their stars earlier (smaller $t_{90}$), compared to their lower stellar mass counterparts at any given halo mass.

\citet{Gallart_2015} propose two distinct evolutionary paths, fast and slow, based on the observed SFHs of Local Group dwarf galaxies. They hypothesise that these evolutionary paths are associated with the characteristic density of the environment where the dwarfs are formed. We find diversity in the SFH of dwarfs as opposed to two distinct categories. However, we find qualitative agreement with their proposal about the origin of the difference in the SFHs.

The underlying physics behind the correlation of stellar assembly and halo concentration (halo assembly) is driven by a combination of various processes; the gas accretion rate onto haloes, the structure of their gravitational potential, and gas heating during and subsequent to the epoch of reionisation. The former two are likely more important at higher halo masses ($\mhalo \gtrsim 10^{11}\Msun$), whereas processes related to cosmic reionisation are thought to be important for less massive haloes ($\mhalo \lesssim 10^9 \Msun$). The details of these processes and the transition scale between various mechanisms are an interesting subject of investigation for future studies, particularly using simulations that explicitly model radiation transport.  

The importance of the correlations we find here between $\mstar, \mhalo, c_{\rm NFW}$, and $t_{90}$ lies in the fact that they link observable properties of dwarf galaxies, i.e. their stellar mass and star formation histories, to their dark matter halo properties which are less readily accessible. 
In particular, we make clear observable predictions for the scatter of the SMHM relation which can be tested with upcoming measurements of properties of large samples of isolated dwarf galaxies in the local Universe, using Euclid, DESI, and the future Rubin and Roman Observatories.

%% file: appendix.tex
\section{Properties of the regions targeted for resimulation}
\label{sec:appendix_region_props}

Key present-day properties of the 25 regions targeted for resimulation, drawn from the parent $L=400\,\cMpc$ cube, are listed in Table \ref{tab:region_details_appendix}. The columns of the table give the region index, the `bands' corresponding to the percentile rank ordering of a Gaussian distribution for, respectively, the enclosed normalised density of the $r=5\,\cMpc$ sphere ($\dfive$) and the number of LMC-mass haloes within the region ($N_{\rm LMC}$), the actual values of $\dfive$ and $N_{\rm LMC}$, and the Cartesian coordinates of the region centre, ($x_{\rm c},y_{\rm c},z_{\rm c}$), in the coordinate system of the parent volume.

\begin{table*}
    \centering
    \caption{Key properties of the regions within the $L=400\,\cMpc$ parent volume that were targeted for resimulation. The columns are as described in the text of Appendix \ref{sec:appendix_region_props}. A visual impression can be obtained by inspection of Fig.~\ref{fig:region_projections}, for which each row corresponds to a band of $\dfive$, and each column to a band of $N_{\rm LMC}$.}
    \begin{tabular}{r|r|r|r|r|r|r|r}
\hline
Region & $\dfive$ & $N_{\rm LMC}$ & $\rm log_{10} \dfive$ & $N_{\rm LMC}$ & $x_{\rm c}$ & $y_{\rm c}$ & $z_{\rm c}$ \\
index & band & band & & & [Mpc] & [Mpc] & [Mpc]\\
\hline
1 &  $-2\sigma$ & $-2\sigma$ & -1.044 & 0 & 22.554 & 61.397 & 182.388 \\
2 &  $-2\sigma$ & $-1\sigma$ & -1.090 & 0 & 285.056 & 380.053 & 5.246 \\
3 &  $-2\sigma$ & $0\sigma$ & -1.075 & 0 & 170.254 & 49.515 & 378.008 \\
4 &  $-2\sigma$ & $+1\sigma$ & -1.084 & 0 & 252.884 & 61.142 & 226.887 \\
5 &  $-2\sigma$ & $+2\sigma$ & -1.053 & 0 & 303.710 & 387.559 & 259.295 \\
6 &  $-1\sigma$ & $-2\sigma$ & -0.851 & 0 & 247.260 & 223.085 & 118.065 \\
7 &  $-1\sigma$ & $-1\sigma$ & -0.829 & 0 & 360.984 & 370.986 & 104.061 \\
8 &  $-1\sigma$ & $0\sigma$ & -0.854 & 0 & 113.541 & 105.269 & 295.703 \\
9 &  $-1\sigma$ & $+1\sigma$ & -0.848 & 1 & 202.167 & 22.568 & 326.103 \\
10 & $-1\sigma$ & $+2\sigma$ & -0.859 & 1 & 327.347 & 244.685 & 199.035 \\
11 & $ 0\sigma$ & $-2\sigma$ & -0.638 & 0 & 206.470 & 289.345 & 337.098 \\
12 & $ 0\sigma$ & $-1\sigma$ & -0.635 & 0 & 164.727 & 370.529 & 389.946 \\
13 & $ 0\sigma$ & $ 0\sigma$ & -0.651 & 1 & 60.793 & 270.998 & 111.084 \\
14 & $ 0\sigma$ & $+1\sigma$ & -0.636 & 2 & 191.408 & 185.739 & 41.606 \\
15 & $ 0\sigma$ & $+2\sigma$ & -0.641 & 4 & 382.208 & 57.656 & 160.808 \\
16 & $+1\sigma$ & $-2\sigma$ & -0.436 & 1 & 211.498 & 78.734 & 381.412 \\
17 & $+1\sigma$ & $-1\sigma$ & -0.432 & 2 & 77.794 & 100.334 & 185.028 \\
18 & $+1\sigma$ & $ 0\sigma$ & -0.445 & 4 & 236.746 & 148.260 & 82.034 \\
19 & $+1\sigma$ & $+1\sigma$ & -0.444 & 6 & 124.015 & 262.135 & 382.086 \\
20 & $+1\sigma$ & $+2\sigma$ & -0.442 & 8 & 356.896 & 181.048 & 185.781 \\
21 & $+2\sigma$ & $-2\sigma$ & -0.279 & 4 & 134.931 & 23.480 & 373.677 \\
22 & $+2\sigma$ & $-1\sigma$ & -0.282 & 6 & 96.567 & 46.163 & 314.870 \\
23 & $+2\sigma$ & $ 0\sigma$ & -0.298 & 8 & 101.094 & 182.843 & 370.797 \\
24 & $+2\sigma$ & $+1\sigma$ & -0.290 & 11 & 211.361 & 265.825 & 313.522 \\
25 & $+2\sigma$ & $+2\sigma$ & -0.292 & 13 & 51.271 & 95.723 & 98.979 \\
\hline
\end{tabular}
\label{tab:region_details_appendix}
\end{table*}

\section{Numerical convergence tests}
\label{appendix:convergence}

In this section we present simple numerical convergence tests of the halo mass function (HMF), the halo concentration - mass relation, and the luminous fraction as a function of halo mass, recovered from the composite halo populations of all 25 simulations at m5 (maroon curves) and m6 (orange curves) resolution. 

\begin{figure}
    \centering
	\includegraphics[width=\columnwidth]{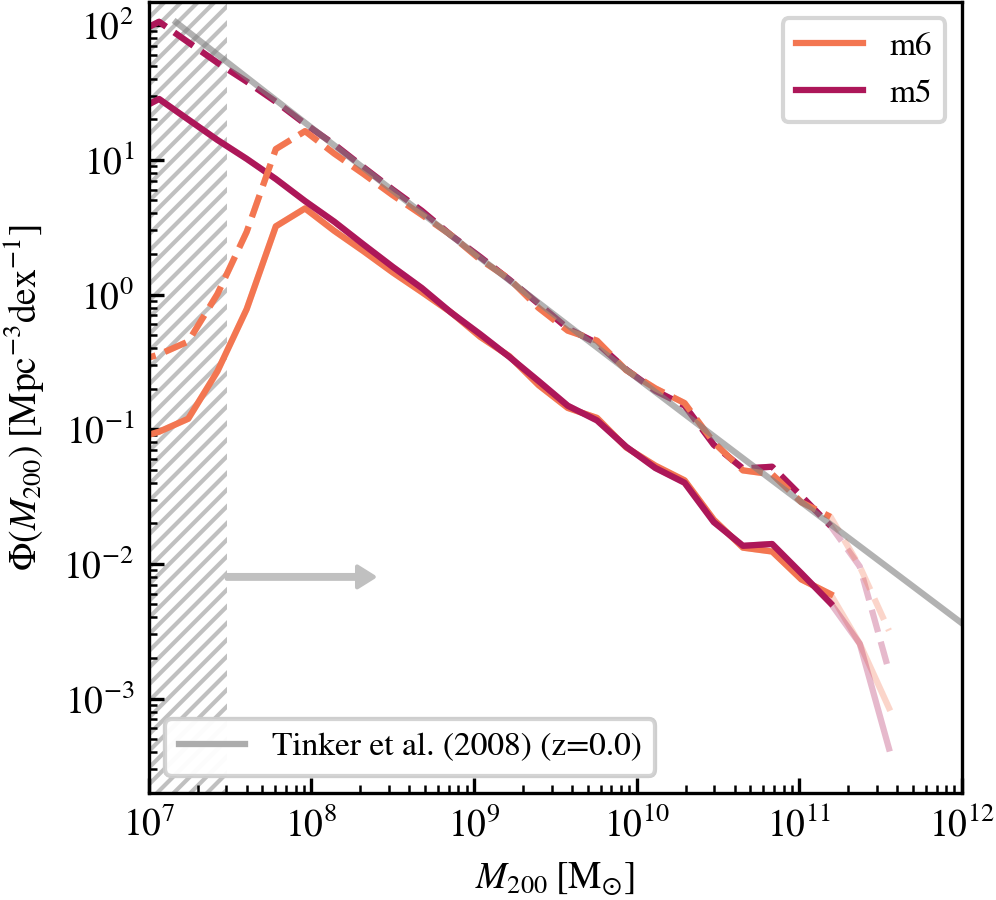}
    \caption{A convergence test of the hydrodynamical HMF with the change in resolution between m6, in orange, and m5, in maroon. The specific resolution details are given in Table \ref{tab:resolution}. Each line combines all 25 regions. At the high-mass end, the lines have lower opacity when the bin contains fewer than 10 haloes.
    For reference, the grey solid line corresponds to the halo mass function from \citet{Tinker_2008}.
    The shaded region indicates 100 m5 resolution DM particles and extends to 100 m6 resolution DM particles indicated by the arrow. When the low mean density of our regions are normalised out, the HMFs shift upwards (dashed lines) to match that of \citet{Tinker_2008} very well.}
    \label{fig:HMF_convergence}
\end{figure}

The convergence of the HMF is shown in Fig.~\ref{fig:HMF_convergence}. As noted in Section \ref{sec:environment:hmf}, the HMFs of the regions are systematically shifted to lower number density at fixed mass relative to the cosmic mean HMF, owing to their relatively low enclosed density. To enable comparison with the cosmic mean HMF of \citet{Tinker_2008}, denoted by the grey curve, the low opacity curves show the HMF recovered when normalising the number density of haloes from each region by the reciprocal of the region's enclosed density ($\dfive$) prior to constructing the composite HMF. This exercise highlights that the renormalised HMFs reproduce the cosmic mean HMF accurately in the mass regime that is well sampled by the low density regions simulated here ($\mhalo \lesssim 10^9\,\Msun$). The figure clearly highlights the excellent convergence of the HMF between the two adopted resolutions: in the mass regime for which haloes are sampled by at least 100 particles at m6 resolution and bins are sampled by at least 10 haloes, the HMFs differ by a maximum of $0.07\,{\rm dex}$ at fixed halo mass.

\begin{figure}
    \centering
	\includegraphics[width=\columnwidth]{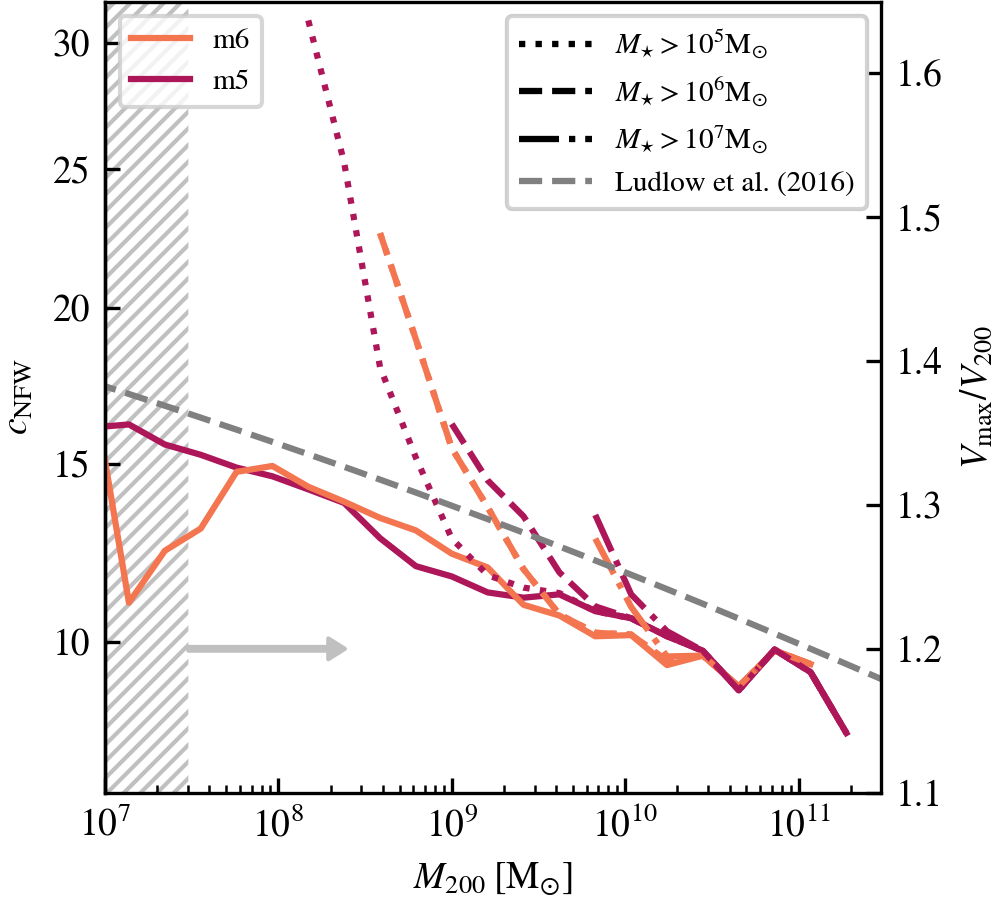}
    \caption{A strong convergence test of the halo mass - concentration relation in the hydrodynamical runs between the m5 resolution, in maroon, and m6, in orange. The NFW concentration parameter is estimated from the ratio of $V_{\rm max}/V_{200}$ from the matched DMO haloes using Eq. \ref{eq:c_conversion}. At the high-mass end, the lines have lower opacity when the bin contains fewer than 10 haloes. Colours and hatching follow the conventions in Fig.~\ref{fig:HMF_convergence}.
    We show corresponding relations for haloes that host a ``luminous'' component, whilst changing the stellar mass threshold for defining luminous: $M_{\star}>10^5\Msun$, $10^6 \Msun$ and $10^7\Msun$ shown as dotted, dashed and dotted-dashed lines, respectively. For the lowest stellar mass threshold, only m5 results are shown since galaxies with $M_{\star}<10^6\Msun$ cannot form in m6 simulations due to its particle mass resolution.  The mass - concentration relation from \citet{Ludlow_2016} is shown as a grey dashed line, for reference.}
    \label{fig:concentration_convergence}
\end{figure}

Fig.~\ref{fig:concentration_convergence} shows the convergence of the halo mass - concentration relation. The relation recovered from the overall halo population (denoted by solid curves) is generally well converged between resolutions, in the mass regime where haloes are sampled by more than 100 particles in both resolutions, differing up to $0.03\,{\rm dex}$. The figure also shows the relation recovered from `luminous' haloes, using different threshold stellar masses to define a luminous halo. \correction{These are $M_{\star}>10^5\Msun$ (dotted curves), $M_{\star}>10^6\Msun$ (dashed curves) and $M_{\star}>10^7\Msun$ (dot-dashed curves).} Using a threshold of $M_{\star}>10^7\Msun$, which corresponds to galaxies sampled by at least 10 and 100 star particles at m6 and m5 resolution, respectively, yields a halo mass - concentration relation that differs by no more than $0.05\,{\rm dex}$ at fixed halo mass. Use of a lower threshold unsurprisingly yields poorer convergence, owing to inadequate particle sampling of the galaxies. 

\begin{figure}
    \centering
    \includegraphics[width=\columnwidth]{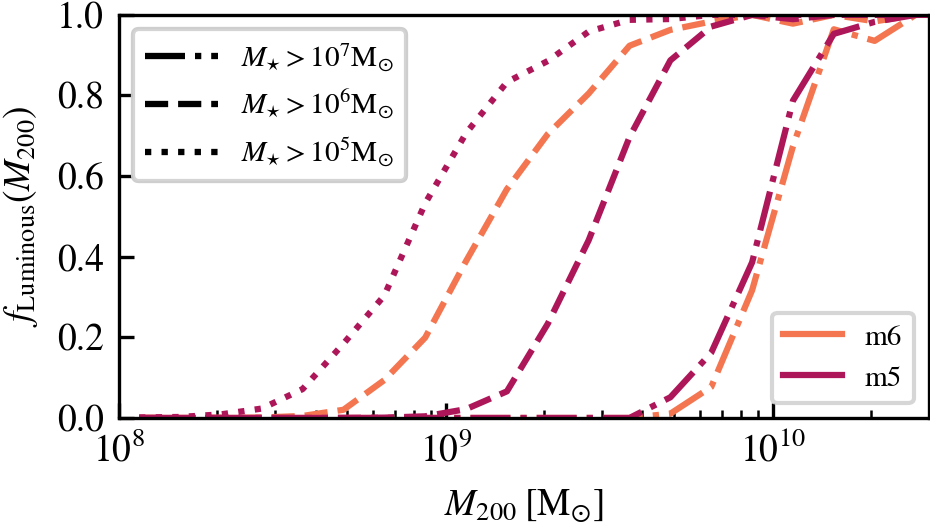}
    \caption{The present-day luminous fraction as a function of halo mass, shown at m5 and m6 resolution in maroon and orange respectively. We explore the effect of changing the stellar mass threshold defining a `luminous' galaxy: we show $M_{\star}>10^5\Msun, 10^6\Msun,$ and $10^7 \Msun$ with dashed, solid, and dashed-dotted lines.}
    \label{fig:flum_conv}
\end{figure}

Fig.~\ref{fig:flum_conv} shows the convergence of luminous fraction as a function of halo mass, at different resolutions (m5 and m6) and different stellar mass thresholds. The dotted lines correspond to a `luminous' threshold of $M_{\star}>10^{5}\Msun$, the dashed lines represent $M_{\star}>10^{6}\Msun$, and the dashed-dotted lines represent $M_{\star}>10^{7}\Msun$. As previously discussed, a threshold of $M_{\star}>10^6\Msun$ is effectively an upper limit for the halo occupation in m6, and similarly $M_{\star}>10^5\Msun$ for m5, formed from one stellar particle. The positions of these curves depend on the choice of stellar mass threshold and reach $100\%$ occupancy at mass scales corresponding to the mass below which the luminous mass-concentration relations diverge from the overall median shown in Fig.~\ref{fig:concentration_convergence}.

For a threshold of $M_{\star}>10^{7}\Msun$, the luminous fraction is well converged between resolutions, differing by less than $11\%$ at fixed halo mass. The convergence for a threshold of $M_{\star}>10^{6}\Msun$ is much poorer, with m6 simulations exhibiting higher luminous fractions differing up to $50\%$ at fixed halo mass. We see a similar poor convergence of the GSMF between m5 and m6 at $M_{\star}<10^7\Msun$  (see, Fig.~\ref{fig:GSMF}). In this regime, galaxies are sampled by fewer than 10 stellar particles in m6, so that the poor convergence is not surprising.

\section{star formation histories as a function of stellar mass}
\label{appendix:avg_t90}
\begin{figure}
    \centering
    \includegraphics[width=\columnwidth]{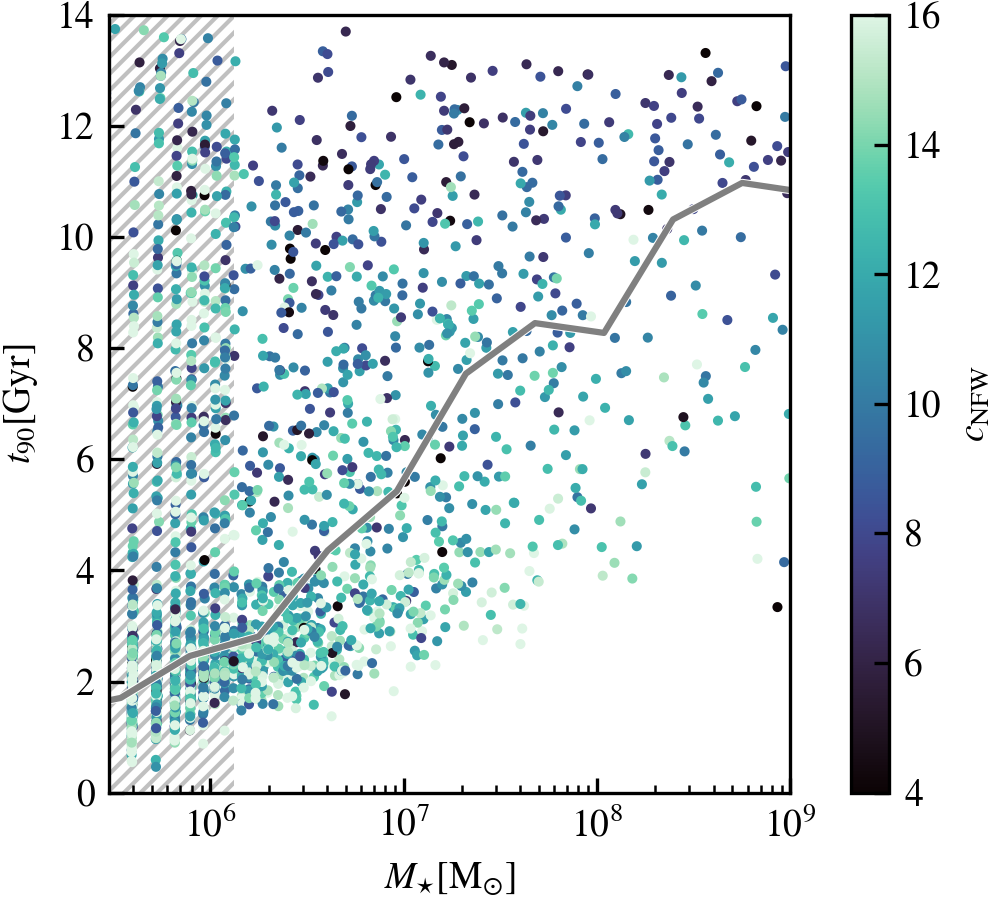}
    \caption{The time it takes for a galaxy to assemble 90\% of its total stellar mass, $t_{90}$, as a function of its present day stellar mass, $M_{\star}$. Each galaxy is coloured by its halo concentration, measured from its DM-only counterpart. The median relation is shown by the grey line.}
    \label{fig:mstar_t90}
\end{figure}

As discussed in Section~\ref{sec:scatter_smhm:tform}, there is an anti-correlation between the scatter about the median stellar mass at fixed halo mass, and duration of star formation quantified by $t_{90}$. This may seem at odds with the expectation that more massive dwarf galaxies exhibit more prolonged SFHs. However, this trend applies only when considering galaxies hosted by haloes of fixed mass. Fig.~\ref{fig:mstar_t90} shows $t_{90}$ as a function of stellar mass for all individual simulated galaxies regardless of their halo mass, with the median trend denoted with a solid curve. The overall trend between stellar mass and the duration of the SFH is positive. Each galaxy in the figure is colour coded by its halo concentration. Consistent with the findings in Section~\ref{sec:scatter_smhm:tform} the scatter about the median correlates (weakly) with halo concentration: at a fixed stellar mass, galaxies with a delayed SFH and hence a greater value of $t_{90}$ are typically hosted by low concentration, and hence later forming, haloes, illustrating how halo concentration drives the scatter in $t_{90}$ at fixed halo mass.

\section{Stellar mass - halo mass relation Fit}
\label{appendix:fit}
\begin{figure}
    \centering
    \includegraphics[width=\columnwidth]{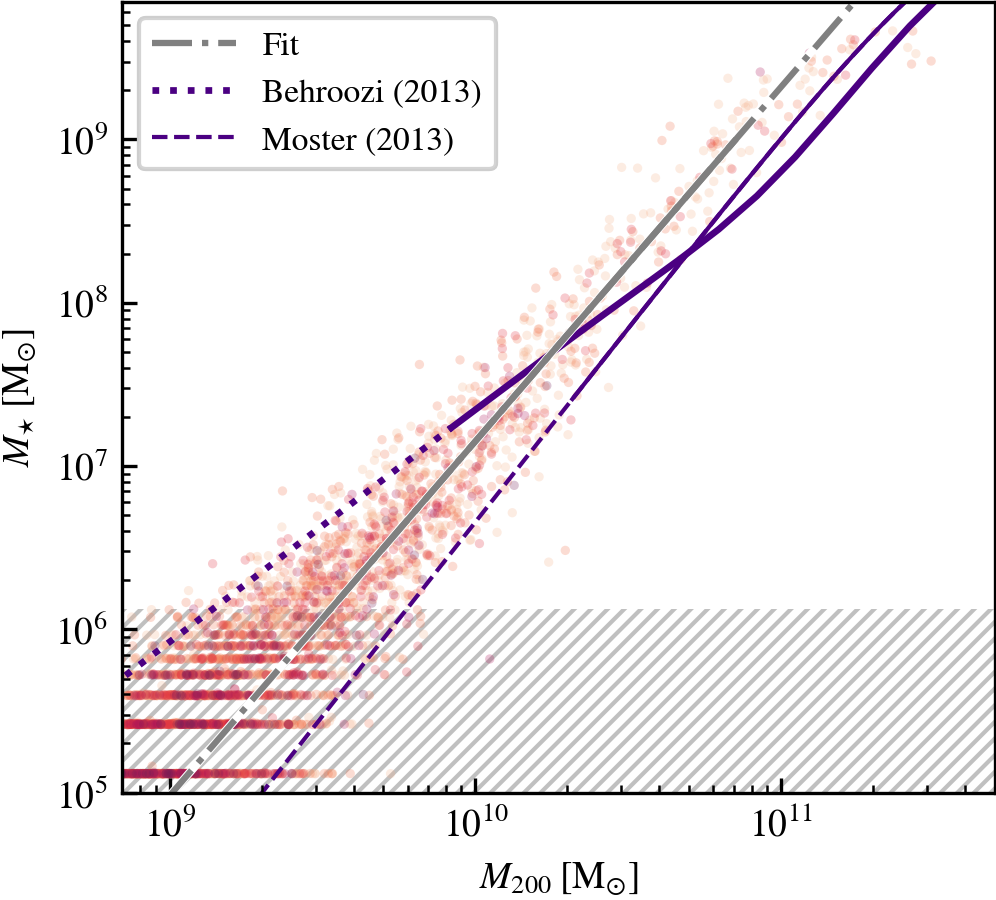}
    \caption{The present-day stellar mass - halo mass scaling relation of central galaxies for the composite population of all 25 regions, at m5 resolution. The grey hatched region indicates the stellar mass regime corresponding to 10 or fewer stellar particles at m5 resolution. The median relation is well fit by a power-law of form $M_{\star} /\Msun = \beta (M_{\rm 200} /10^{10}\Msun)^{\alpha}$ where $(\alpha,\beta)=(2.17,1.42\times 10^7)$, fitted to the range $M_{\star}= [10^6-10^9] \Msun$ and shown by the grey line. Indigo curves show the SMHM relations of the semi-empirical models of \citet{Behroozi_2013} and \citet{Moster_2013}, \correction{which transition to solid lines in the mass range to which they are fit}. All lines are solid for the mass scale they are fit to.}
    \label{fig:smhm_fit}
\end{figure}

Fig. \ref{fig:smhm_fit} shows the present-day SMHM relation of the composite population of field galaxies formed from all 25 simulated regions at m5 resolution. The median SMHM relation can be represented by a power-law of the form $M_{\star} /\Msun = \beta (M_{\rm 200} /10^{10}\Msun)^{\alpha}$ where $(\alpha,\beta)=(2.17,1.42\times 10^7)$, fitted to the range $M_{\star}= [10^6-10^9] \Msun$, as is shown by the grey line in Fig. \ref{fig:smhm_fit}. The semi-empirical models of \citet{Behroozi_2013} and \citet{Moster_2013} are shown in indigo for comparison.

